\documentclass[aps,prb,twocolumn]{revtex4}

\usepackage{graphicx}%
\usepackage{dcolumn}
\usepackage{amsmath}

\makeatletter
\def\btt#1{\texttt{\@backslashchar#1}}%
\DeclareRobustCommand\bblash{\btt{\@backslashchar}}%
\makeatother

\begin{document}

\title[Short Title]{ Critical charge instability  on  verge of the  Mott transition\\
 and the origin of quantum  protection  in  high-$T_c$ cuprates}
 
 \date{\today}

\begin{abstract}
The concept of topological excitations and the related  ground state degeneracy are employed to
establish an effective theory of the superconducting state evolving from the Mott insulator
for high-$T_c$ cuprates. The theory includes the effects of the relevant energy scales with the emphasis
 on the  Coulomb interaction $U$ governed by the electromagnetic U(1) compact group. The results are
obtained for the layered $t-t'-t_\perp-U-J$ system of strongly correlated  electrons relevant for cuprates.
 Casting  the  Coulomb interaction  in terms of composite-fermions via the gauge flux attachment facility,
 we show that instanton  events in the  Matsubara ``imaginary time",
labeled by a topological winding  numbers, are essential configurations of the phase field dual to the charge.
 This provides a non-perturbative concept of the topological quantization and  displays
 the 	significance of  discrete topological sectors in the theory governed by the global characteristics
 of the phase field.  We  show  that for topologically ordered states these  quantum numbers 
 play the role of an  order parameter in a way similar to the Landau order parameter  
for conventionally ordered states. In analogy to the usual phase transitions that are 
 characterized by a sudden change of the symmetry, the topological phase transitions are governed
by a discontinuous change of the topological numbers signalled by the divergence of 
 the zero-temperature topological susceptibility. This defines a novel type quantum criticality
 ruled by topologically conserved  numbers  rather than the Landau principle of the symmetry breaking.
We show that in the limit of strong correlations  topological charge 
is linked to the average electronic filling number
and the topological susceptibility to the electronic compressibility of the system. 
We exploit the impact of these  nontrivial U(1) instanton phase field configurations  
for the cuprate phase diagram which displays the ``hidden" quantum critical point covered by the
 superconducting lobe in addition to a sharp crossover between a compressible
normal ``strange metal"  state and a region characterized by a vanishing
compressibility, which marks the Mott insulator.
Finally, we argue that the existence of robust  quantum numbers explains
the stability against  small perturbation of the system  and attributes to the
topological ``quantum protectorate" as observed in strongly correlated systems.
\end{abstract}
\pacs{74.20.Fg, 74.72.-h, 71.10.Pm}	

\author{T. K. Kope\'{c}}
\affiliation{
Institute for Low Temperature and Structure Research,\\
Polish Academy of Sciences,\\
POB 1410, 50-950 Wroclaw 2, Poland}
\maketitle
%
\section{Introduction}
%
Superconductivity represents a remarkable phenomenon where quantum coherence
effects appear at macroscopic scale.\cite{bcs}
The superconducting (SC) properties,
especially the perfect diamagnetism, are microscopic manifestations of the spontaneous
breakdown of one of the fundamental symmetries of matter, namely the U(1)
gauge symmetry.\cite{schrieffer}
For a superconductor this introduces a state  with no definite particle
number, but with a definite conjugate variable identifiable with the  {\it phase}. 
The discovery of the cuprate superconductors,\cite{mb} which is believed to emerge due
to the strong electron-electron interactions, has
sparked a widespread interest in physics which goes beyond
the traditional Fermi-liquid framework usually employed for understanding the
effect of interactions in metals.\cite{anderson1}
It has been  recognized that the superconductivity occurs in the region of the doped Mott
insulator near the Mott transition,
so that the whole microscopic foundations on which BCS theory
was built on cannot be accommodated to explain cuprate superconductivity.
The conventional BCS theory identifies SC order
as an instability of the Fermi sea. That assumes the existence of a normal
Landau-Fermi-liquid\cite{flt} which forms well above the critical temperature  $T_c$,
which eventually turns into a superconductor
below $T_c$ due to  residual attraction among low-lying quasiparticles.
 In cuprates, in turn, there is  overwhelming evidence that superconductivity 
does not appear as an instability of a Landau-Fermi liquid.  The reason is that
the fermion quasiparticle
does not seem to reflect the character of the measured low energy eigenstates. 
In this sense, the electron (or electron-like quasiparticle) may no longer be the appropriate
way to think of elementary excitations.
It is intriguing to conjecture
that in the strongly correlated systems the ubiquitous competitions between the variety
of possible ground states govern the essential physics - the formation of a
highly degenerate state  seems to open the way to transformation 
into alternative stable electronic configurations.
Strong correlations that suppress electron motion, may transform the system into
a kind of a  unstable state which will be very sensible to   charge and/or spin
ordering. In this  critical--like state the superconductivity might emerge as a competition
between different ground states.
 Indeed, in cuprates there is clear evidence for the existence of a special
doping point ``hidden" near optimal doping below
the SC dome where superconductivity is
most robust.\cite{bob,dicastro} Such a behavior suggests that  this point could be a  quantum critical point
(QCP) while  the associated critical fluctuations might be responsible for the 
unconventional normal state properties.\cite{qcp} However, 
it is unclear whether this QCP is ``truly critical" in the sense
that it is characterized by universality and scaling.
For example, if  excitations at QCP also carry the
electrical current, then a resistivity linear in temperature, as is observed in cuprates,
 obtains only if the dynamical exponent is  unphysically negative.\cite{chamon}
The resemblance to a conventional QCP is also hampered by the lack of any clear signature of
thermodynamic critical behavior.
Usually, a QCP would be the end-point of a critical line below which an  ordered phase takes place 
and it could be made manifest below the superconducting dome.
Experiments appear to exclude any broken  symmetry 
around this point  although  a sharp change in transport properties is 
observed. \cite{transport}
Unfortunately, our understanding of  the underlying orders in cuprates is far from 
being satisfactory and identifying the nature of the putative
QCP	is an open question.
A possible way out from this difficulty would be if the degrees of freedom that rule
QCP's    are different  from  the  energy degrees 
of freedom that govern the stable phases the critical point separates. Thus these
``non-conventional" degrees of freedom could provide critical fluctuations
beyond those of the order parameter fluctuations 
usually included in the standard Ginsburg-Landau-Wilson 
(LGW) description.\cite{lgw1,wilson}
Within the LGW approach the order parameter fluctuations
are implemented   by constructing models which mimic the {\it low} energy
properties of solids - a procedure which  relies on the separability of the high and low energy
scales.  Under this proviso 
 the high energy variables can be removed out yielding
an effective Hamiltonian (as exemplified e.g.
by the $t-J$ model\cite{rice}) which describes the relevant
low energy and long-wavelength physics. 
This procedure is often implemented via the 
projective transformation, which results in 
removing  of high-energy degrees of freedom
and replacing them with kinematical constraints.
In such approaches, the high energy scale associated with the
charge gap is argued to be irrelevant, hence the focus exclusively 
on the spin sector to characterise the Mott insulator.
However, the charge transfer nature
of the cuprates plays an essential role in the doped
systems\cite{zan} so that  with discarding charge degrees of freedom an important part of
the physics may be lost.
However, there is also mounting   experimental evidence  which put in question the validity
of the various projection schemes.
For example, it has been found that above any temperature
associated with ordering in both electron and
hole-doped cuprates a charge gap of order $2eV$ is present
in the optical conductivity and a rapid reshuffling of spectral weight with  hole doping.\cite{cooper,uchida}
 Angle--resolved photoemission
spectroscopy (ARPES) also reveals a similar charge gap.\cite{armitage}
Surprisingly, when superconductivity emerges, the low and high
energy degrees of freedom are still coupled. It has been shown that
changes in the optical conductivity occur at energies
$3eV$ which exceeds by two orders of  magnitude the maximum of the superconducting gap.\cite{rubh}
But the high energy scale  involved is hard to understand unless it is assumed that
the spectral weight is transferred from both the lower and
the upper Hubbard bands, thus beyond the range
of applicability of the pure 	 $t-J$ model.
The  intimate connectedness
between the low and high energy degrees of freedom in
doped Mott insulators was firmly appreciated  and termed  {\it mottness}.\cite{motness}
Due to the intrinsic  mixing of high and low energy degrees of freedom
no low energy reduction is possible in a conventional sense, so that the doped Mott insulators are 
inherently asymptotically slaved.\cite{asympt}
In a similar spirit
a detour from the strict projection program was recently proposed in a form of
the ``gossamer" superconductor \cite{laughlin}
recognizing the role of the expensive, double--occupancy charge
configurations.

While  spontaneous symmetry breaking  has become one of the main
guiding principles in physics,\cite{pwa2} there are  other signatures  in a physical system
that are associated  with the {\it topological} effects. 
These 	are instrumental for a full understanding of the physics 
and lead to a host of rather unexpected and exotic phenomena, which
are in general of a nonperturbative nature.
For example, the fundamental character
of a vector potential is evident in the Aharonov--Bohm (AB) effect,\cite{ab}
 where the topology  of the U(1)  group  is essential:
when an electron is transported in a magnetic field
around a closed loop, it acquires a phase that is equal to the
magnetic flux through the surface spanned by the electron path.
Strongly correlated electronic
system are no exception in this regard.
In particular, the fractional quantum Hall  (FQH) effect\cite{fqhe1,fqhe2}  is  the prominent representative. 
Here, the striking fact is that FQH systems may contain many different
phases at zero temperature which have the same symmetry. Thus different
states cannot be distinguished by symmetries and the  Landau
symmetry-breaking principle fails
because also  topological characteristics of the configurational space come into play.
In the most interesting cases configurational spaces are not simply connected. There are
space--time configurations of quantum fields  which cannot be
continuously deformed one into another.
Further, an adiabatic motion along a noncontractible
closed path in a configurational
space leads to a geometric (Berry) phase\cite{berry}  acquired by the wave function. 
In condensed matter systems with large
numbers of mutually interacting particles, the subject of Berry's phases becomes a key issue.
For example, a Berry phase distinguishes between integer and half-integer spin chains
and results in different ground states and excitations.\cite{haldane}
The other examples are the possible geometric phases' effects on statistical 
transmutation\cite{trans} that
can be achieved by a ``flux attachment"  which now becomes
a very powerful theoretical method.\cite{trans2,wilczek2}
In many cases the topological character of the quantum field is captured by single integer,
called the topological charge, or winding number of the field which classify topological excitations.
These are found by integrating the so--called Chern-Simons terms which enter the Lagrangians of the theory.
What is common to all the above issues  is the appearance of  {\it gauge} fields to
characterize various  interactions: field configurations which differ by a gauge transformation are to be regard
as physically the same.

In the present work we argue that the important properties of cuprates are 
controlled by the large Mott gap and consider the representation of strongly
correlated electrons as  fermions with attached ``flux tubes".
This introduces a conjugate U(1) phase variable, which acquires
dynamic significance from the electron--electron interaction.
This means that an electron is not a quasiparticle (in the Landau sense), but
has a composite nature governed by the electromagnetic gauge group. 
Furthermore, we recognize  the non-trivial topology of the electromagnetic U(1) group
by observing  that the fundamental group $\pi_1(U(1)) =Z$ is given by a set of integers. Therefore
the elementary excitations in a strongly correlated system always carry $2\pi$-kinks of the phase
field  characterized by the topological winding number\cite{sakola} - a quantized U(1) topological charge. 
Due to the nontrivial first homotopy group
$\pi_1$  the configuration space of the quantum phase fields is multiply connected, so that
inequivalent paths in the ``imaginary--time" (paths that are not deformable to one
another) naturally emerge. Hence, the U(1) gauge field
gives rise to a ``topological interaction", which is felt by the electrons and  can be separated from ordinary 
dynamical ones lumping them into particle ``statistics".
To facilitate this task we employ the functional
integral formulation of the theory that encompasses all of these topological  possibilities: one has to
perform the functional integral over fields defined on different topologically equivalent classes
i.e,  with different winding numbers. From a canonical (operator)  point of view, however,
the different  topological sectors seem to give rise to completely different Hilbert
spaces and the resultant field operators
would satisfy quite complicated nonlocal commutation relations.
The fact that a prospective theory of electronic states in strongly correlated electron systems must
give up on either standard fermion commutation relations or standard particle conservation laws 
has already been  pointed out.\cite{meinders}
Furthermore, we exploit the impact of these  topological excitations  for the  phase diagram of cuprates
(with its various crossovers and transition lines)
and show that they can induce its
unusual feature: a ``hidden" quantum critical point of the  type that  results from
the topological ground-state degeneracy.
It can be probed by the topological susceptibility
as a robust, nonperturbative property that is related to the physical quantity of interest, namely, 
the diverging charge compressibility.
It also provides natural  description of the Mott state where
the system is said to be incompressible when there is a gap in the
chemical potential as a function of the electron density.
This topological underpinning establishes the source of quantum protection
as a collective state of the quantum field,
whose   excitations pertain to  the whole system. Therefore, macroscopic behaviour is mostly 
determined by topological conservation
laws which  does not arise  just out of a
symmetry of the theory (as ``conventional" conservation laws based on Noether's theorem) 
but it is a consequence of the connectedness, i.e. topology 
of the phase  space,  related to the topological properties of the associated gauge  group manifold.

The organization of this paper is as follows.
In Section II  we introduce the electronic model for cuprates which 
captures the  layered structure and epitomizes
the hierarchy of relevant energy scales, with the largest set up by the Coulomb
interaction. In Section III we describe the details of the
flux attachment transformation which results in   the representation of strongly correlated
electrons as  fermions plus attached U(1) gauge ``flux tubes"
that leads to composite particle picture. Section IV is devoted to the basic concepts
of the  algebraic topology  (homotopy groups)
that  becomes instrumental for  the Feynman's path integral formulation of quantum statistics.
This is followed by Sections V and VI where the 
fermionic part of the theory is elaborated in terms of the momentum dependent 
``$d$-wave" spin--gap and microscopic phase stiffnesses.
Here, the most important results are summarized in the effective bosonic  model
written with the help of the collective phase variables.
This enables us to  study the  superconductivity as  the condensation of the ``flux-tubes" 
from the electron composite  which is presented in
Section VII. We elucidate there  the  role of doping  for the superconducting 
order to occur and the key role played by  the topological degeneracy. 
In the subsequent  Section VIII
the  topological susceptibility  is used to probe the change of the topological order.
Here, we show that  the  topological susceptibility
is related to the charge compressibility that  diverges at the degeneracy
point at zero-temperature and defines a novel type of topological  quantum criticality,
beyond the paradigm of the symmetry breaking.
In Section IX we present calculated phase diagrams for cuprates displaying, 
beyond the conventional ordered states, regions  that are related to the
change of the topological order. Section X is dedicated to the discussion of the
robustness of the ground states  in cuprates   and its source in the topological conservation laws.
Finally, in Section XI we conclude with a summary of the results, while
the Appendixces collect material that is related to the technical part of the work.

\section{Electronic model for cuprates}
%
The Hubbard model is viewed as the generic model for interacting
electrons in the narrow-band and strongly correlated
systems\cite{hubb} that captures the physics of Mott transition.\cite{mott}
The existence of the Mott insulator in the cuprates' parent compounds implies  that
a viable theory of high--temperature superconductivity
{\it must} explicitly incorporate  the Mott-Hubbard gap for charge transfer.\cite{zan}
While it is believed that the basic pairing mechanism in cuprates arises from the antifferomagnetic (AF) exchange
correlations,\cite{ex} it is apparent the charge fluctuations also play an essential role in doped systems.
Hence the  Coulomb charge fluctuations  associated with double occupancy of a site are
controlled by the parameter $U$ in the Hubbard model, which also determines
the strength of the AF exchange coupling $J$.
Therefore  energy scale of the charge fluctuation is characterized by the Mott gap, which
is by far larger than the energy scale of magnetic fluctuations.
Although the $t-J$ model is usually viewed as the  $U\to \infty$ limit
of the single band Hubbard model, the one--particle spectra of the two models
differ considerably.\cite{dagotto}
As already mentioned, this limitation of the
$t-J$ model comes from having projected out the doubly
occupied states originally contained in the Hubbard
model. To explore the more flexible
arrangements between $J$ and $U$ than encoded either in the Hubbard
or $t-J$ models  we employ in the present paper a generalized
 $t-t'-U-J$ model for the CuO$_2$ plane in high$-T_c$
superconducting. 

In this way we retain basic features of both models:  the charge
fluctuations present in the Hubbard model (but removed from  the constrained
$t-J$ model) and the robust superconducting correlations described
by the exchange interaction $J$.
Furthermore, we incorporate besides the  direct nearest neighbor  hopping $t$
also the next nearest neighbor $t'$  hopping parameter, which importance
in cuprates has been emphasized.\cite{pavarini}
A good deal of the existing literature on the cuprates
invokes  model Hamiltonians based only on the properties of a single CuO layer.
Obviously, the interlayer structure cannot be ignored:\cite{leggett}
the measured critical temperature $T_c$ is strongly dependent
on the interlayer structure.\cite{Kuzemskaya} Therefore, three dimensional
(3D) coupling of planes must play an important
role in the onset of superconductivity,\cite{ml1,ml2} which have to be
incorporated by means of the interlayer couplings and the associated $c$-axis
dispersion effects in modeling of the cuprates,\cite{bansil} 

Summing up, we consider an effective one--band electronic
Hamiltonian on a tetragonal lattice that emphasizes strong anisotropy and
the presence of a layered CuO$_2$ stacking sequence in cuprates:
${\cal H}= {\cal H}_t+{\cal H}_J+{\cal H}_U$, where
\begin{eqnarray}
&&{\cal H}_t=\sum_{\alpha\ell}
\left[ -\sum_{\langle {\bf r}{\bf r}'\rangle}
 (t+\delta_{{\bf r,r'}}\mu)c^{\dagger }_{{\alpha}\ell}({\bf r})
c_{\alpha \ell}({\bf r}')
\right.
\nonumber\\
&&+
\left.
\sum_{{\langle \langle{\bf r}{\bf r}'\rangle\rangle}}
 t'c^{\dagger }_{{\alpha}\ell}({\bf r})
c_{\alpha \ell}({\bf r}')
 -
\sum_{{\bf r}{\bf r}'}
t_\perp({\bf r}{\bf r}')
 c^{\dagger }_{{\alpha}\ell}({\bf r})
c_{\alpha \ell+1}({\bf r}')\right],
\nonumber\\
&&{\cal H}_J=\sum_\ell\sum_{{\langle {\bf r}{\bf r}'\rangle}}
{J}\left[{\bf S}_\ell{({\bf r})}
\cdot{\bf S}_\ell{({\bf r}')}
-\frac{1}{4}{n}_\ell{({\bf r})}{n}_\ell{({\bf r}')}\right],
\nonumber\\
&&{\cal H}_U=\sum_{\ell\bf r}
Un_{\uparrow\ell} ({\bf r}) n_{\downarrow\ell}({\bf r}).
\label{mainham}
\end{eqnarray}
Here $\langle {\bf r},{\bf r}'\rangle$ and 
$\langle\langle {\bf r},{\bf r}'\rangle\rangle$
identify  summations
over the nearest-neighbor and next--nearest--neighbor
sites labelled by $1\le {\bf r}\le N$ within the CuO  plane, respectively,
while $1\le\ell\le N_\perp$ labels copper-oxide layers.
Subsequently, $t$ and $t'$ are the {\it bare}  hopping integrals,
while $t_\perp$ stands for the interlayer coupling.
The operator $c_{\alpha\ell}^\dagger({\bf r})(c_{\alpha\ell}({\bf r}))$
creates (annihilates) an electron with spin $\alpha$ at the lattice site $({\bf r},\ell)$
and ${n}_\ell({{\bf r}})= n_{\uparrow\ell} ({\bf r})+n_{\downarrow\ell}({\bf r})$
stand for  number operators, where  $ {n}_{\alpha\ell}({{\bf r}})= c^\dagger_{\alpha\ell}({\bf r})
c_{\alpha\ell}({\bf r})$ and $\mu$ is the chemical potential.
Furthermore,
\begin{equation}
 S^a_{\ell}({\bf r})=\sum_{\alpha\beta}c^\dagger_{\alpha\ell}({\bf r})
\sigma_{\alpha\beta}^a c_{\beta\ell}({\bf r})
\end{equation}
denotes the vector spin operator ($a=x,y,z$) with ${\sigma}^a_{\alpha\beta}$ being
  the Pauli matrices.
Finally, $U$ is the on--site repulsion Coulomb energy  and $J$ the AF exchange. 
	Owing the lattice arrangement
the full electronic dispersion is given by
\begin{eqnarray}
\epsilon({\bf k},k_z)=\epsilon_\|({\bf k})+\epsilon_\perp({\bf k},k_z),
\end{eqnarray}
where the in-plane contribution reads
\begin{eqnarray}
 \epsilon_\|({\bf k})&=&-2t\left[\cos(ak_x)+\cos(ak_y)\right]
 \nonumber\\
&+&4t'\cos(ak_x)\cos(ak_y)
 \end{eqnarray}
with $t'>0$. Furthermore, the $c-$axis dispersion is given by
\begin{eqnarray}
&&\epsilon_\perp({\bf k},k_z)=2t_\perp({\bf k})\cos(ck_z),
\nonumber\\
&&t_\perp({\bf k})=t_\perp\left[\cos(ak_x)-\cos(ak_y)\right]^2
\end{eqnarray}
as predicted on the basis of  band calculations.\cite{ander}
%
\section{Electron as a composite object}
We now provide the representation of interacting
electrons as  fermions plus attached ``flux tubes".\cite{trans2}
This leads to a picture of composite particles which 
are void of the mutual  interactions among fermions: 
the electron-electron  Coulomb interaction  will be transformed into
the action of  U(1) gauge (phase) fields governed by the effective  kinetic term
of ``free" quantum rotors.
%
\subsection{Fermionic action}
%
The partition function  for the system governed by the Hamiltonian in Eq.(\ref{mainham})
can be represented as a path integral using fermionic coherent-states. 
Introducing Grassmann fields $c_{\alpha\ell}({\bf r}\tau),{\bar c}_{\alpha\ell}({\bf r}\tau)$
that depend on the ``imaginary time" $0\le\tau\le \beta\equiv 1/k_BT$,
with $T$ being the temperature, we write the path integral for the statistical sum ${\cal Z}$ as
\begin{eqnarray}
{\cal Z}=\int\left[{\cal D}\bar{c}  {\cal D}\bar{c}
\right]e^{-{\cal S}[\bar{c},c]}
\end{eqnarray}
with the fermionic action 
\begin{eqnarray}
{\cal S}[\bar{c},c]=\int_0^\beta d\tau\left[\sum_{\alpha{\bf r}\ell}
 \bar{c}_{\alpha\ell}({\bf r}\tau)\partial_\tau{c}_{\alpha\ell}({\bf r}\tau)+{\cal H}(\tau)\right].
 \end{eqnarray}
The Hubbard term in Eq.(\ref{mainham}) we write  in
a SU(2) invariant way as
\begin{equation}
{\cal H}_U(\tau)=U\sum_{{\bf r}\ell}\{({1}/{4}){n_\ell}^2({{\bf r}}\tau)
-\left[{\bf \Omega}_\ell({\bf r}\tau)\cdot{\bf S}_\ell({\bf r}\tau)\right]^2\}
\label{huu}
\end{equation}
singling out the charge-U(1)  and  spin-SU(2)/U(1)   sectors, where the unit vector
${\bf \Omega}_\ell({\bf r}\tau)$ labels
varying in space-time  spin quantization axis. \cite{weng}
The spin--rotation invariance one can made explicit
by performing angular integration over  ${\bf\Omega}({\bf r}\tau)$
at each site and time. In the following we fix our attention on the
U(1) invariant {\it charge }sector leaving aside  possible magnetic
orderings such as antiferromagnetism. 
Although sometimes concurrent magnetic transitions occur
with the Mott transition, the mechanism of the Mott transition is primarily independent
of the symmetry breaking of spins.
Thus  we stress our
primary interest in cuprates due to their superconducting properties
and the fact that the superconductivity (resulting from condensation of  charge)
should not be viewed as inextricably connected with the  quantum antiferromagnetism.
A clear support for this point comes from the observation of the spectral weight transfer
through the superconducting  transition in cuprates,\cite{rubh}
 which cannot be explained by invoking
the antiferromagnetic order: the spectral weight transfer persists well above the Neel
temperature  and at the  doping level where antiferromagnetism is absent.

%
\subsection{Gauge flux attachment transformation}
To proceed, we  employ the Hubbard-Stratonovich transformation
to decouple the Coulomb  term in Eq.(\ref{huu}) with the help of the fluctuating  imaginary
	electrochemical potential   $iV_\ell({\bf r}\tau)$
conjugate to the  charge number $n_\ell({\bf r}\tau)$.
Furthermore,  we write the field  $V_\ell({\bf r}\tau)$ as a sum of
a static $V_{0\ell}({\bf r})$ and periodic function\cite{kopec}
\begin{eqnarray}
\tilde{V}_\ell({\bf r}\tau)&\equiv&\tilde{V}_\ell({\bf r}\tau+\beta)
\nonumber\\
V({\bf r}\tau)&=&V_0({\bf r})+\tilde{V}({\bf r}\tau),
\end{eqnarray}
where using  Fourier series
\begin{eqnarray}
\tilde{V}({\bf r}\tau)=({1}/{\beta})\sum_{n=1}^\infty
[\tilde{V}({\bf r}\omega_n)e^{i\omega_n\tau}+c.c.]
\end{eqnarray}
with $\omega_n=2\pi n/\beta$ ($n=0,\pm1,\pm2$)
being the (Bose) Matsubara frequencies.
Now, we introduce the {\it phase } (or ``flux") field ${\phi}_\ell({\bf r}\tau)$
via the Faraday--type relation
\begin{equation}
\dot{\phi}_\ell({\bf r}\tau)\equiv\frac{\partial\phi_\ell({\bf r}\tau)}
{\partial\tau}=\tilde{V}_\ell({\bf r}\tau),
\label{jos}
\end{equation}
to remove the imaginary term
\begin{equation}
 i\int_0^\beta d\tau\dot{\phi}_\ell({\bf r}\tau) n_\ell({\bf r}\tau)
\equiv i\int_0^\beta d\tau\tilde{V}_\ell({\bf r}\tau)n_\ell({\bf r}\tau)
\end{equation}
 for all the Fourier modes
of the $V_\ell({\bf r}\tau)$ field, except for  the zero frequency
by performing the local gauge transformation to the {\it new} fermionic
 variables $f_{\alpha\ell}({\bf r}\tau)$:
\begin{eqnarray}
c_{\alpha\ell}({\bf r}\tau)=
\exp\left[ i\int_0^\tau d\tau' \tilde{V}_\ell({\bf r}\tau')\right]
f_{\alpha\ell}({\bf r}\tau).
\label{compo}
\end{eqnarray}
Thus as a result of Coulomb correlations the electron acquires a phase shift similar to that
in the 	electric (i.e. scalar) AB effect.\cite{ab} The expression in
Eq.(\ref{compo}) means that an electron 
has a composite  nature made of the fermionic part $f_{\alpha\ell}({\bf r}\tau)$
with the attached ``flux" (or  AB phase) $\exp[i\phi_\ell({\bf r}\tau)]$.
Here, the quantity ${\cal C}_0({\bf r}\tau)$  defined by
\begin{equation}
{\cal C}_{\ell 0}({\bf r}\tau)\equiv\dot{\phi}_\ell({\bf r}\tau)=\tilde{V}_\ell({\bf r}\tau),
\label{cs}
\end{equation}
is the one dimensional (temporal) component Chern-Simons term\cite{chern}
that makes the minimal coupling with the fermion density field.
Since the abelian Chern-Simons term is just as a total (time) derivative.
the integral of it becomes simply converted by into a ``surface" integral, sensitive only to the 
global properties of the U(1) gauge field along a ``imaginary time" path that
starts  at imaginary time $\tau=0$ and ends at $\tau=\beta$.
Thus  the paths can be divided
into topologically distinct classes, characterized by a winding
number  defined as the net number of times the world
line wraps around the system in the ``imaginary time" direction.
As we shall see in the next Section, our considerations  related to the ``imaginary time"
boundary conditions
can be formalized using  homotopically non-trivial gauge transformations for which the
strength of the phase shift must be quantized, so that the gauge change of the Chern-Simons
term will be an integral multiple of $2\pi$.
%
\section{Quantum states  on multiply connected  spaces}
%
%
\subsection{Homotopy theory}
The algebraic topology and precisely the concept of homotopy groups\cite{mermin}  provides 
 the necessary background by making reference to the topological structure of the group manifold,
 let  us say  $M$. The $n-$th homotopy group $\pi_n(M)$ is a group of equivalent classes of 
loops that can be smoothly deformed into each other without leaving $M$, where $n$
refers to the dimensionality of the loops in question.
The already mentioned  AB effect  is  of topological nature, since mappings from
the electron's configuration space to the gauge group  constitute the
non-trivial  homotopy group $\pi_1(U(1)) =Z$, where the elements of $Z$ are integers and represent
winding numbers, i.e, U(1)  topological charges.	
We are precisely facing a similar situation in a strongly correlated system since the electromagnetic 
U(1) gauge group governs the charge sector. This becomes apparent in an electrodynamic
dual description of the charge in terms of the phase field, see Eq.(\ref{jos}),
having it's roots in the quantum mechanical complementarity of phase and number. 
In the ``imaginary time" evolution  of the phase, two field configurations lie 
in the same connected component of configurational space if they can be continuously 
deformed into each other. There is a natural equivalence relation between these paths
 called homotopy: two paths are equivalent (i.e. belong to the same class) if they can 
be ``smoothly deformed" into each other. The classes are labeled by the winding numbers 
 and are endowed with a group structure by appropriately defining the composition
 of two mappings.\cite{mermin} 
The rule of thumb is that if the homotopy group  is trivial, then there cannot be any
topological field configurations in the underlying  theory.
%
\begin{figure}
\begin{center}
\includegraphics[width=5cm]{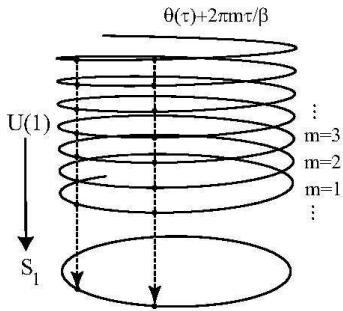}
\end{center}
 \caption{Schematic representation of the mapping from the real line $R$ 
(the covering group of $U(1)$) to the circle $S_1\sim U(1)$
 which is locally invertible provided the topological sector with the winding integer  number $m$ is chosen.
 The map is defined by a continuous function $f(\theta)$ on $[0,2\pi]$, where $f(0)=0$ and
 continuity of the map requires that $f(2\pi)=2\pi m$, so that the homotopy classes of $\pi_1(S_1)$
 are labelled by integers.
   }\label{fig1}
\end{figure} 

\subsection{Homotopy classes and the path integral}
If we work in the Feynman's path integral  formulation of the quantum statistics\cite{schulman}
then the statistical sum ${\cal Z}$  takes a form,
in which  homotopically distinct paths have to be summed
according to  various possibilities for inequivalent
quantizations (superselection sectors). Specializing to the
case of  the U(1) group one obtains
\begin{equation}
{\cal Z}=\sum_{m\in \pi_1(U(1))}\rho(m)Z(m)
\label{zm1}
\end{equation}
Here, $m\in Z$ labels equivalence classes of homotopically connected paths and  $\rho(m)$
marks the ``statistical " weight which is related to a homotopy class.
Furthermore, the partial sum ${\cal Z}(m)$ within the $m$-th topological sector is given  by the usual path integral 
\begin{equation}
{\cal Z}(m)=\sum_{m}\int\left[ {\cal D}\phi\right]_m\left[{\cal D}\bar{f}  {\cal D}{f}
\right]e^{-{\cal S}[\phi,\bar{f},f]}
\label{zm}
\end{equation}
with the integration  restricted to the $m$-th homotopy class. Furthermore,
the weights factors $\rho(m)$ form unitary irreducible representations 
of the homotopy group $\pi_1(U(1))$, so that the conditions for the weights  are
\begin{eqnarray}
&&|\rho(m)|=1
\nonumber\\
&&\rho(m_1)\rho(m_2)=\rho(m_1 \star m_2).
\label{ht2}
\end{eqnarray}
In Eq.(\ref{ht2}) $m_1$ and $m_2$ label   the homotopy classes of two the paths with common end points,
while $m_1 \star m_2$ label the homotopy class of the path obtained by joining the two.
In particular, the weight factor, which furnishes the representations of $Z$
takes the form $\rho(m)=e^{i\theta m}$, where $\theta\in[0, 2\pi)$ is the ``statistical angle" 
 parameter.\cite{gross} From the canonical point of view,
if the configuration space of the  system is not simply connected, as for the U(1) group,
then the quantization prescription becomes ambiguous since paths belonging to different
 homotopy classes can get the extra relative Berry phase factor  acquired by the wave
 function and the $\theta$
factor  represents exactly this quantization ambiguity. Moreover,  the $\theta$ term
cannot be traced in a perturbation theory because it has no affect upon the equations of motion.
Therefore, in performing the integration in Eq.(\ref{zm}) one should take phase configurations
that satisfy  the boundary condition 
$\phi_\ell({\bf r}\beta)-\phi_\ell({\bf r}0)=2\pi m_\ell({\bf r})$
and $m_\ell({\bf r})=0,\pm 1,\pm 2,\dots$. For the sake of convenience it is
desirable to ``untwist" the boundary condition by setting (see, Fig.\ref{fig1})
\begin{equation}
\phi_\ell({\bf r}\tau)\longrightarrow\phi_\ell({\bf r}\tau)\equiv\varphi_\ell({\bf r}\tau)+
\frac{2\pi\tau}{\beta}m_\ell({\bf r}).
\end{equation}
Summing over all the phases $\phi_\ell({\bf r}\tau)$ amounts to integrating over
the $\beta$--periodic field $\varphi_\ell({\bf r}\tau)$ ($\varphi_\ell({\bf r}0)=\varphi_\ell({\bf r}\beta)$)
and to sum over the set of integer winding numbers $\{m_\ell({\bf r})\}$. 
The integral over  the {\it static} (zero frequency)  part of the
fluctuating  electrochemical potential
$\left\langle V_0({\bf r}) \right\rangle$ we calculate by the
saddle point method to give
\begin{eqnarray}
\frac{2}{iU}\left\langle V_0({\bf r}) \right\rangle&=&
\left\langle\sum_\alpha\bar{f}_{\alpha}({\bf r}\tau)
f_{\alpha}({\bf r}\tau)\right \rangle
+\frac{2\mu}{U}.
\label{ex3}
\end{eqnarray}
Explicitly, following the
prescription given in Eqs.(\ref{zm1}) and (\ref{zm}) we obtain for the statistical sum
\begin{widetext}
\begin{eqnarray}
{\cal Z}&=&\sum_{ \{m_\ell({\bf r})\}}\int_0^{2\pi}\prod_{{\bf r}\ell}d\varphi_{0\ell}({\bf r})
\int_{\phi_\ell({\bf r}0)=\varphi_{0\ell}({\bf r})}^{\phi_\ell({\bf r}\beta)
=\varphi_{\ell 0}({\bf r})+2\pi m_\ell({\bf r})}
\prod_{{\bf r}\ell\tau}{d}\varphi_\ell({\bf r}\tau)\int\left[{\cal D}\bar{f}  {\cal D}{f}
\right] e^{-{\cal S}[{\varphi},m,\bar{f},f]},
\label{explicit1}
\end{eqnarray}
with the action involving the topological Chern-Simons term and the statistical angle parameter
\begin{eqnarray}
{\cal S}[{\varphi},m,\bar{f},f]&=&\sum_{ \ell}
\int_0^\beta d\tau\left\{
\frac{1}{U}\sum_{\bf r}\left[ 
\frac{\partial\varphi_\ell({\bf r}\tau)}{\partial\tau}+\frac{2\pi}{\beta} m_\ell({\bf r})\right]^2
+\frac{2\mu}{U}\sum_{ {\bf r}}
\frac{1}{i}\left[ 
\frac{\partial\varphi_\ell({\bf r}\tau)}{\partial\tau}+\frac{2\pi}{\beta} m_\ell({\bf r})\right]
\right.
\nonumber\\
&+&\left.{\cal H}\left[ {\phi},\bar{f},f  \right]\right\}.
\label{explicit2}
\end{eqnarray}
Here, ${\cal H}\left[ {\phi},\bar{f},f  \right]$ is  the effective Hamiltonian that is void of the
Coulomb interaction
\begin{eqnarray}
{\cal H}\left[ {\phi},\bar{f},f  \right] &=& \sum_{ \ell}
\int_0^\beta d\tau\left\{
\sum_{{ \bf r}\alpha}
\bar{f}_{\alpha\ell}({\bf r}\tau)\partial_\tau
 f_{\alpha\ell}({\bf r}\tau)
-\bar{\mu}\sum_{{ \bf r}\alpha}
\bar{f}_{\alpha\ell}({\bf r}\tau)
 f_{\alpha\ell}({\bf r}\tau)\right.
\nonumber\\
&-&\sum_{\langle {\bf r},{\bf r}'\rangle} te^{-i[\phi_\ell({\bf r}\tau)
-\phi_\ell({\bf r}'\tau)]}
 \sum_\alpha\bar{f}_{{\alpha}\ell}({\bf r}\tau)
f_{\alpha \ell}({\bf r}'\tau)
+
\sum_{{\langle\langle  {\bf r},{\bf r}'\rangle\rangle}}t'e^{-i[\phi_\ell({\bf r}\tau)
-\phi_\ell({\bf r}'\tau)]}
\sum_\alpha \bar{f}_{{\alpha}\ell}({\bf r}\tau)
f_{\alpha \ell}({\bf r}'\tau)
\nonumber\\
&+&\left.
\sum_{\langle {\bf r},{\bf r}'\rangle }
t_\perp({\bf r}{\bf r}')e^{-i[\phi_\ell({\bf r}\tau)
-\phi_{\ell+1}({\bf r}'\tau)]}
\sum_\alpha\bar{f}_{{\alpha}\ell}({\bf r}\tau)
f_{\alpha \ell+1}({\bf r}'\tau)-J\sum_{\langle{\bf r r'\rangle}}
\bar{\cal B}_\ell( {\bf r}\tau,{\bf r}'\tau)
{\cal B}_\ell( {\bf r}\tau,{\bf r}'\tau)\right\},
\label{explicit}
\end{eqnarray}
\end{widetext}
where $\bar{\mu}=\mu-n_f{U}/{2}$ is the shifted chemical potential, while
\begin{equation}
n_f=\sum_\alpha\langle \bar{f}_{\alpha}({\bf r}\tau)
f_{\alpha\ell}({\bf r}\tau)\rangle
\label{nfdef}
\end{equation}
is the occupation number for the fermionic part of the electron composite. 
In Eq.(\ref{explicit}), while writing the term that governs AF interaction
we made use of the following representation
\begin{eqnarray}
\bar{\cal B}_\ell( {\bf r}\tau,{\bf r}'\tau)
=\frac{1}{\sqrt{2}}[\bar{f}_{\uparrow\ell}({\bf r}\tau)
\bar{f}_{\downarrow\ell}( {\bf r}'\tau)
-\bar{f}_{\downarrow\ell}( {\bf r}\tau)
\bar{f}_{\uparrow\ell}( {\bf r}'\tau)]
\end{eqnarray}
which is just the singlet--pair (valence bond) operator\cite{bond}
emerging from the decomposition
\begin{eqnarray}
&&J\sum_\ell\sum_{\langle{\bf r r'\rangle}}\left[{\bf S}_\ell{({\bf r}\tau)}
\cdot{\bf S}_\ell{({\bf r}'\tau)}
-\frac{1}{4}{n}_\ell{({\bf r}\tau)}{n}_\ell{({\bf r}'\tau)}\right]
\nonumber\\
=&&-J\sum_\ell\sum_{\langle{\bf r r'\rangle}}
\bar{\cal B}_\ell( {\bf r}\tau,{\bf r}'\tau)
{\cal B}_\ell( {\bf r}\tau,{\bf r}'\tau).
\label{rvb}
\end{eqnarray}
It is  obvious that a quasiparticle description (of any kind) makes sense
only when the constituent objects are weakly interacting.
The chief merit of the transformation in Eq.(\ref{compo}) is that we have managed 
to cast the strongly correlated problem into a system of
{\it weakly interacting} $f$-fermions with residual interaction given by  $J$,
submerged in the bath of strongly fluctuating U(1) gauge potentials
 (on the high  energy scale set by $U$)
minimally coupled to $f$-fermions via ``dynamical Peierls" phase factors.
It is clear that the action of these phase factors ``frustrates" the motion of the
fermionic subsystem. However, as we demonstrate in the following, it is only when charge 
fluctuations become {\it phase coherent}
the frustration of the kinetic energy is released. 
From Eq.(\ref{explicit2}) we can read off the explicit expression for the statistical weight 
\begin{eqnarray}
\rho(m)
&=&\exp\left[i\cdot \frac{2\mu}{U}\cdot 2\pi m_\ell({\bf r}) \right]
\label{rhom}
\end{eqnarray}
so that the statistical angle reads
\begin{equation}
\frac{\theta}{2\pi}\equiv\frac{2\mu}{U}.
\label{statangle}
\end{equation}
The phase factor in Eq.(\ref{rhom}), being a topological quantity is closely related to the concept of the geometric 
Berry phases. To explicate this w we write explicitly  the composite  structure
of the physical electron field using Eq.(\ref{compo}):
\begin{eqnarray}
c_{\alpha\ell}({\bf r}\tau)=
\exp\left[ i\int_0^\tau d\tau' \tilde{V}_{D\ell}({\bf r}\tau')\right]e^{i\gamma_{\ell B}({\bf r}\tau)}
f_{\alpha\ell}({\bf r}\tau).
\label{compo2}
\end{eqnarray}
The first term in the exponential in Eq.(\ref{compo2}) is the usual dynamical phase factor where 
$\dot{\phi}_\ell({\bf r}\tau)=\tilde{V}_{D\ell}({\bf r}\tau)$ and
${\phi}_\ell({\bf r}\beta)=\phi_\ell({\bf r}0)$ .
The second one, in turn,  is the {\it non-integrable} Berry phase factor:\cite{berry}
$\gamma_{\ell B}({\bf r}\tau)=2\pi\tau m_\ell({\bf r})/\beta$,
where $m_\ell({\bf r})$ marks the integer U(1) topological  number.
%
\subsection{Topological ground state degeneracy}
\begin{figure}
\begin{center}
\includegraphics[width=7cm]{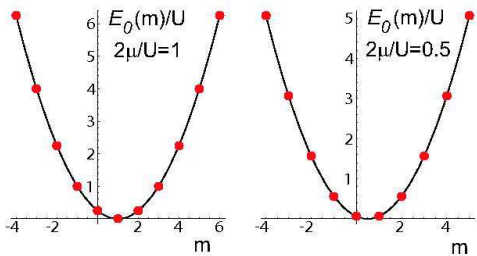}
\end{center}
 \caption{Energy levels of the quantum  Hamiltonian, Eq.(\ref{rot}). Half--odd integer values of
the ``statistical" parameter  $2\mu/U$ yield
 ground state degeneracy.}
\label{fig2}
\end{figure}

%
It is known that the ground state degeneracy may arise from broken symmetries.
Here, we argue that the ground state
degeneracy of the  charge states in strongly correlated system states is a reflection
of the topological properties of the system.
The existence of topological features implies that the quantum eigenstates
are not single-values under the continuation  of the parameters in the Hamiltonian.
Consider the ``free" part of the action in Eq.(\ref{explicit2}) describing the dynamics of the U(1)gauge 
field
\begin{eqnarray}
{\cal S}_0[{\phi}]&=&\sum_{ \ell}
\int_0^\beta d\tau\left\{
\frac{1}{U}\sum_{\bf r}\left[ 
\frac{\partial\phi_\ell({\bf r}\tau)}{\partial\tau}\right]^2\right.
\nonumber\\
&+&\left.\frac{2\mu}{U}\sum_{ {\bf r}}
\frac{1}{i}
\frac{\partial\phi_\ell({\bf r}\tau)}{\partial\tau}
\right\}.
\label{freeaction}
\end{eqnarray}
This is the action of a particle  moving in a plane around  ``magnetic flux "
$\Phi/\Phi_0\equiv 2\mu/U$.
The Hamiltonian corresponding to this action is simply
\begin{eqnarray}
{\cal H}_0[{\phi}]&=&
\frac{U}{4}\sum_{{\bf r}\ell}\left[ 
\frac{\partial}{\partial\phi_\ell({\bf r})}-\frac{2\mu}{U}\right]^2,
\label{rot}
\end{eqnarray}
with the eigenenergies given by
\begin{eqnarray}
{\cal E}_0(m)&=&
\frac{U}{4}\sum_{{\bf r}\ell}\left[ m_\ell({\bf r})-\frac{2\mu}{U}\right]^2.
\label{enrot}
\end{eqnarray}
The energy in Eq.(\ref{enrot}) can be interpreted as the square of the kinetic angular
momentum of a set  of quantum rotors divided by the  ``moment of inertia" $I=2/U$
and the allowed angular momenta are uniformly displaced from integers by $2\mu/U$, which may be
any real number. The energy spectrum is labeled by integers $m_\ell({\bf r})$ and for integer
$2\mu/U$ the ground state is non-degenerate. However, for the half-odd integer  $2\mu/U$ the 
ground state is doubly degenerate, see Fig.\ref{fig2}. In this case  the destructive interference
between even and  odd topological sectors is responsible for the ground state degeneracy.
As we shall  see in the following, this degeneracy will be crucial
in explaining the occurrence of superconductivity and anomalous properties on the verge
of Mott transition,
where  the correspondence between the filling
factor and the ground state degeneracy will also be established.
%
%
\section{Bare and dressed band parameters}
%
With the help of the  angle-resolved photoemission spectroscopy
one gets a direct access to the density of low-energy electronic excited
states in the momentum-energy space of Cu-O planes in cuprates.
Obviously, all the interactions of the electrons   are
encapsulated in ARPES data, but these  are still difficult to evaluate.
Since the underlying band structure of the {\it bare }electrons is
{\it a priori} unknown, one way to think about these interactions is to consider
simply the electronic excitations as quasiparticles which  are characterized by 
{\it effective} electronic  parameters. Consequently, the tight-binding interpolation
 of the electronic structure is  often used for fitting the experimental  ARPES  data for cuprates.
In this type of  analysis a conceptually simpler band theory   is used to reveal
how the strongly correlated electronic effects  can be taken  into account via the influence 
of the electronic band parameters. In the context of the present work in order to establish 
a link between the ``bare" band parameters of the high--energy model in Eq.(\ref{mainham})
and the ``dressed" one  of the low energy models we have to perform the averaging over the
 fluctuation of the U(1) gauge phase fields according to
\begin{eqnarray}
\int\left[ {\cal D}\phi\right]\left[{\cal D}\bar{f}  {\cal D}{f}
\right]e^{-{\cal S}[\phi,\bar{f},f]}
=\int\left[{\cal D}\bar{f}  {\cal D}{f}
\right]e^{-{\cal S}_{LE}[\bar{f},f]},
\label{}
\end{eqnarray}
where the low energy action ${\cal S}_{LE}$ is given by 
\begin{eqnarray}
{\cal S}_{LE}[\bar{f},f]=-\ln
\int\left[ {\cal D}\phi\right]e^{-{\cal S}[\phi,\bar{f},f]}.	
\label{x1}
\end{eqnarray}
Performing the cumulant expansion in  Eq.(\ref{x1}) we can deduce in the lowest 
order the corresponding
low energy fermionic Hamiltonian that has to be compared with the original one in Eq.(\ref{mainham}):
\begin{eqnarray}
&&{\cal H}_{LE} = \sum_{ \ell}
\left\{
 -\sum_{\langle {\bf r},{\bf r}'\rangle,\alpha}( t^\star+\delta_{{\bf r},{\bf r}'}\bar{\mu})
 {f}^\dagger_{{\alpha}\ell}({\bf r})
f_{\alpha \ell}({\bf r}') \right.
\nonumber\\
&&
+
\sum_{{\langle\langle  {\bf r},{\bf r}'\rangle\rangle,\alpha}}t'^\star
{f}^\dagger_{{\alpha}\ell}({\bf r})
f_{\alpha \ell}({\bf r}')-J\sum_{\langle{\bf r r'\rangle}}
{\cal B}^\dagger_\ell( {\bf r},{\bf r}')
{\cal B}_\ell( {\bf r},{\bf r}')
\nonumber\\
&&+\left.
\sum_{\langle {\bf r},{\bf r}'\rangle,\alpha }
t_\perp^\star({\bf r}{\bf r}')
{f}^\dagger_{{\alpha}\ell}({\bf r})
f_{\alpha \ell+1}({\bf r}')\right\},
\end{eqnarray}
where the dressed parameters encapsulating the effect of Coulomb interaction  are given by
\begin{eqnarray}
&&t^\star=t \langle e^{-i[\phi_\ell({\bf r}\tau)
-\phi_\ell({\bf r}'\tau)]}\rangle\delta_{|{\bf r}-{\bf r}'|,1st}
\nonumber\\
&&t'^\star=t'\langle e^{-i[\phi_\ell({\bf r}\tau)
-\phi_\ell({\bf r}'\tau)]}\rangle\delta_{|{\bf r}-{\bf r}'|,2nd}
\nonumber\\
&& t_\perp^\star({\bf r}{\bf r}')= t_\perp({\bf r}{\bf r}')\langle e^{-i[\phi_\ell({\bf r}\tau)
-\phi_{\ell+1}({\bf r}'\tau)]}\rangle
\label{baredress}
\end{eqnarray}	
and $\langle\dots\rangle$ refers to the averaging over the U(1) phase field.
\begin{equation}
\langle\dots\rangle\equiv\frac{\int [{\cal D}\phi]\dots e^{-S_0[\phi]}}
{\int [{\cal D}\phi] e^{-{\cal S}_0[\phi]}}
\end{equation}
with,  the strongly fluctuating kinetic part (on the energy scale $U$) ${\cal S}_0[\phi]$
given by Eq.(\ref{freeaction}).
On average, the effect of this renormalization due to the presence of the 
phase--phase correlation functions is the effective mass enhancement of the
 carriers as a result of  the band narrowing, so that the ``dressed" band 
parameters  $t^\star_X$ (where $t_X=t,t',t_\perp$) are  used to match electronic
 spectra using the  {\it low}--energy scale $t-J$ model.\cite{spectra}
Typically, in cuprates $t^\star\sim 0.5$eV, $t'^\star/t^\star\sim 0.15-0.35$ and
$t^\star_\perp$ is of order of magnitude smaller then the in--plane hopping parameters.\cite{ander}
%
\section{Pseudogap and phase stiffnesses}
%
%
\subsection{RVB pairs: single-particle ``$d$-wave" gap }
A routine Hubbard-Stratonovich decoupling is applied to the resonating
valence bond (RVB) term in Eq.(\ref{rvb}) to give
\begin{eqnarray}
&&\exp\left[ -J\int_0^\beta d\tau\sum_\ell\sum_{\langle{\bf r r'\rangle}}
\bar{\cal B}_\ell( {\bf r}\tau,{\bf r}'\tau)
{\cal B}_\ell( {\bf r}\tau,{\bf r}'\tau)\right]=
\nonumber\\
&&=\int\left[{\cal D}\Delta^\star  {\cal D}
\Delta\right]
\int	 e^{-{\cal S}\left[\bar{f},f,\Delta^\star,\Delta\right]},
\end{eqnarray}
where $\Delta({\bf r}\tau;{\bf r'}\tau)$ is the complex pair field that is non-local in space.
Furthermore, the corresponding effective action is of the form

\begin{widetext}
\begin{eqnarray}
{\cal S}\left[\bar{f},f,\Delta^\star,\Delta\right]&=&
\frac{1}{J}
\sum_\ell\sum_{\langle {\bf r}{\bf r}'\rangle}\int_0^\beta
 d\tau
|\Delta_\ell({\bf r}\tau,{\bf r}'\tau)|^2-
\sum_\ell
\sum_{\langle {\bf r}{\bf r}'\rangle}\int_0^\beta d\tau
\left\{\frac{\Delta_\ell({\bf r}\tau,{\bf r}'\tau)}
{\sqrt{2}}\left[f_{\uparrow}({\bf r}\tau)
f_{\downarrow}( {\bf r}'\tau)
-f_{\downarrow}( {\bf r}\tau)
f_{\uparrow}( {\bf r}'\tau)
\right]\right.
\nonumber\\
&+&
\left.
\frac{\Delta^\star_\ell({\bf r}\tau,{\bf r}'\tau)}
{\sqrt{2}}\left[\bar{f}_{\uparrow}({\bf r}\tau)
\bar{f}_{\downarrow}( {\bf r}'\tau)
-\bar{f}_{\downarrow}( {\bf r}\tau)
\bar{f}_{\uparrow}( {\bf r}'\tau)
\right] \right\}.
\label{finaction}
\end{eqnarray}
\end{widetext}
A saddle-point method  can applied to the action in Eq.(\ref{finaction}) 
to give  the  self-consistency equation for the momentum--dependent gap
parameter $|\Delta({\bf k})|$ belonging to the Cu-O plane. 
Assuming that $\Delta({\bf k})$ is not changing  along the $c$--direction, we can drop the
layer index for this quantity.
Out of the many possible mean-filed  translationally invariant solutions in the RVB
theory the ``$\pi$-flux" phase is selected here
because of its  relation to the ``$d$-wave" symmetry
of the pseudogap in cuprates. It is governed by the equation
\begin{eqnarray}
&&1
=\frac{J}{N}\sum_{\bf k}\frac{\eta^2({\bf k})}{2{E}({\bf k})}
\tanh\left[
\frac{\beta{E}({\bf k})}{2} \right],
\label{rvbsol}
\end{eqnarray}
where, $\eta({\bf k})=\cos(k_xa)
-\cos(k_ya)	$ with the quasiparticle spectrum of the fermionic part of the electron composite,
\begin{equation}
 {E}^2({\bf k})=[\epsilon_\|^\star({\bf k})-\bar{\mu}]^2
+|\Delta({\bf k})|^2  
\end{equation}
and  
\begin{equation}
\Delta({\bf k})=\Delta[\cos(ak_x)
-\cos(ak_y)].
\end{equation}
The gap parameter is a  quantity with the short--range property,
$\lim_{|{\bf r}-{\bf r'}|\to\infty}\Delta({\bf r}-{\bf r}')=0$, essentially tied
to local correlations on neighboring sites. 
As we see in the following,  the presence of the ``$d$-wave" pair function $\Delta({\bf k})$
is {\it not} a signature of the superconducting state - it merely marks the region of
non-vanishing phase stiffness.
However, the presence of $\Delta({\bf k})$ should be visible 
e.g., in ARPES spectra that picture  the momentum-space occupation
and therefore can detect the dispersion ${E}({\bf k})$
with a  gap for  single particle excitations.\cite{armitage}
%
\subsection{Microscopic phases stiffnesses}
%
We can take advantage of the effective
fermionic  action in Eq.(\ref{finaction}) which  is now
quadratic in the  Grassmann field variables that can
be integrated out without any difficulty yielding
a fermionic determinant:
\begin{equation}
Z=\int\left[{\cal D}\phi \right]e^{-{\cal S}_0[\phi]+{\rm Tr}\ln \hat{G}^{-1}},
\label{intout}
\end{equation}
where
\begin{equation}
\hat{G}^{-1}=\widehat{G}^{-1}_{\it o}-{{\bf T}}=
\left(1-{{\bf T}}\widehat{G}_{\it o}\right)\widehat{G}^{-1}_{\it o}
\end{equation}
is the effective propagator of the theory, while
%
%
\begin{eqnarray}
&&{\bf T}\equiv\left[{\bf T}\right]_{\ell\ell'}
({\bf r}\tau,{\bf r}'\tau')=
\nonumber\\
&&\left\{
te^{-i[\phi_{\ell}({\bf r}\tau)
-\phi_{\ell}({\bf r}'\tau)]\hat{\sigma_3}}
\delta_{\ell\ell'}\delta_{|{\bf r}-{\bf r}'|,1st}\hat{\sigma}_3\delta(\tau-\tau')
\right.
\nonumber\\
&-&t'e^{-i[\phi_{\ell}({\bf r}\tau)
-\phi_{\ell}({\bf r}'\tau)]\hat{\sigma_3}}
\delta_{\ell\ell'}\delta_{|{\bf r}-{\bf r}'|,2nd}
\nonumber\\
&+&\left.t_\perp({\bf r}-{\bf r}') e^{-i[\phi_{\ell}({\bf r}\tau)
-\phi_{\ell'}({\bf r}'\tau)]\hat{\sigma_3}}
\delta_{|\ell-\ell'|,1}\right\}\hat{\sigma}_3\delta(\tau-\tau')
\nonumber\\
&\equiv&\hat{{\cal T}}+\hat{\cal T}' +\hat{\cal T}_\perp
\end{eqnarray}
%
is the matrix composed of the hopping integrals and phase factors.
Furthermore,
\begin{eqnarray}
\left[\widehat{G}^{-1}_{\it o}\right]_{\ell}
({\bf r}\tau,{\bf r}'\tau')
&=&\left[\left(
\frac{\partial}{\partial\tau}
-\hat{\sigma}_3\mu 
\right)
\delta_{{\bf r},{\bf r}'}+
\frac{\Delta_{\ell}({\bf r}\tau,{\bf r}'\tau)}
{\sqrt{2}}\hat{\sigma}_+
\right.\nonumber\\
&+&\left.
\frac{\Delta_{\ell}^\star({\bf r}\tau,{\bf r}'\tau)}{\sqrt{2}}
\hat{\sigma}_-\right]
\delta_{\ell\ell'}\delta(\tau-\tau')
\end{eqnarray}
stands for  the inverse of the ``free" fermion propagator containing  the gap $\Delta$ field.
Here,
\begin{eqnarray}
\hat{\sigma}_+&=&\frac{1}{2}\left(\hat{\sigma}_1+i\hat{\sigma}_2\right),
\nonumber\\
	\hat{\sigma}_-&=&\frac{1}{2}\left(\hat{\sigma}_1-i\hat{\sigma}_2\right),
\end{eqnarray}
where $\hat{\sigma}_a$  $(a=1,2,3)$ are  the Pauli matrices acting in the
Nambu spinor space, so that
\begin{eqnarray}
&& \hat{G}_{\it o}({\bf r}\tau{\bf r}'\tau')=\left[
\begin{array}{cc}
{\cal G}({\bf r}\tau{\bf r}'\tau')
 & {\cal F}({\bf r}\tau{\bf r}'\tau')  \\
{\cal F}^\star({\bf r}\tau{\bf r}'\tau') &-{\cal G}({\bf r}'\tau'{\bf r}\tau)
\end{array}
\right],
\end{eqnarray}
Using the  self--consistency solution, Eq.(\ref{rvbsol}),
and Fourier transforming to the frequency and momentum domain one obtains
\begin{eqnarray}
&& \hat{G}_{\it o}
({\bf k}\nu_n)=\left(
\begin{array}{c}
\frac{-i\nu_n +\mu}
{\nu^2_n+\mu^2
+|\Delta({\bf k})|^2}
, \frac{|\Delta({\bf k})|}{\nu^2_n
+\mu^2+|\Delta({\bf k})|^2}\\
 \frac{|\Delta({\bf k})|}{\nu^2_n
+\mu^2+|\Delta({\bf k})|^2},
 \frac{-i\nu_n -\mu}{\nu^2_n
+\mu^2+|\Delta({\bf k})|^2}
\end{array}
\right),
\end{eqnarray}
where $\nu_n=(2n+1)\pi/\beta$ are the (Fermi) Matsubara frequencies,
$n=0,\pm 1,\pm2,\dots$. 
Now, expanding the trace of the logarithm in Eq.(\ref{intout}) 
 we obtain up to the second order in the hopping matrix elements
\begin{eqnarray}
{\rm Tr}\ln{\hat{G}}^{-1}
&&={\rm Tr}\hat{ G}^{-1}_{\it o}
-\frac{1}{2}{\rm Tr} \left(\hat{G}_{\it o}\hat{\cal T }\right)^2
-{\rm Tr} \left(\hat{ G}_{\it o}\hat{\cal T }'\right)
\nonumber\\
&&
-\frac{1}{2}{\rm Tr} \left(\hat{G}_{\it o}\hat{\cal T }_\perp\right)^2
+\dots.
\label{expand}
\end{eqnarray}
Finally, by performing summations over frequencies and momenta that are implicitly
assumed in the trace operation  in Eq.(\ref{expand})
we obtain an effective action expressed in the U(1) phase fields
\begin{eqnarray}
{\cal S}_{ph}[\phi]&=&\sum_{\ell}\int_0^\beta d\tau \left\{\sum_{ {\bf r}}
\left[ 
\frac{1}{U}\dot{\phi}^2_\ell({\bf r}\tau)
+\frac{2\mu}{U}
\frac{1}{i}
\dot{\phi}_\ell({\bf r}\tau)\right]\right.
\nonumber\\
&-&\sum_{{\langle{\bf r}{\bf r}'\rangle}}
{\cal J}_\|(\Delta)\cos\left[2\phi_\ell({\bf r}\tau)
-2\phi_\ell({\bf r}'\tau)\right]
\nonumber\\
&-&\sum_{{\langle\langle{\bf r}{\bf r}'\rangle\rangle}}
{\cal J}'_\|(\Delta)\cos\left[\phi_\ell({\bf r}\tau)
-\phi_\ell({\bf r}'\tau)\right]
\nonumber\\
&-&\left.\sum_{{\bf r}}
{\cal J}_\perp(\Delta)\cos\left[2\phi_\ell({\bf r}\tau)
-2\phi_{\ell+1}({\bf r}\tau)\right]\right\},
\label{phasemodel}
\end{eqnarray}
In our case
$U \gg {\cal J},{\cal J}',{\cal J}_\perp$ so that the microscopic phase stiffnesses can
be regarded as  residual interactions to the dominant  kinetic term
 of the phase model in Eq.(\ref{phasemodel}) .
This justifies the retention of only the lowest order non-vanishing terms in
the electron hopping $t$ and $t'$ in Eq.(\ref{expand}) since for the ensuing  microscopic
phase stiffnesses one has, eg.$\frac{{\cal J}(\Delta)}{U}\sim t/U$ and
 $\frac{{\cal J}'(\Delta)}{U}\sim t'/U$,
respectively. 
All the stiffnesses in Eq.(\ref{phasemodel}) (see also Eq.(\ref{stiff}) in Appendix A)
rest on the single-particle gap due to the  in-plane  {\it momentum space }  pairing among
fermionic parts of the electron composites   governed by  the  AF exchange $J$:
when $\Delta({\bf k})=0$ all the phase stiffnesses  collapse.
While ${\cal J}_\|$ and ${\cal J}_\perp$ depend on the square
of the corresponding  hopping elements, the stiffness ${\cal J}'_\|$
is different: it depends {\it linearly} on $t'$
and governs the process of correlated   {\it particle-hole} motion.
Collective  pair  transfer events  are costly for large $U$,
so that  excitonic coherent charge transfer dominates the in-plane  charge motion.\cite{kopec}
The inter-plane stiffness ${\cal J}_\perp$  is essential, however, in establishing 
bulk superconductivity via the Josephson-like interplanar  coupling.
%
\section{From the Mott insulator to superconductor}
%
%
\subsection{ODLRO vs. charge frustration}
%
Because of the composite nature of the electron field in strongly
correlated system the occurrence   of superconductivity
requires {\it both} the condensation of the
of fermion pairs described by  $\bar{f}_{\downarrow}( {\bf r}\tau)$ ,
$\bar{f}_{\uparrow}( {\bf r}\tau)$
as well as the phase coherence which follows form the condensation of ``flux tubes" $e^{i\phi_\ell({\bf r}\tau)}$
attached to the $f$-fermions. 
Thus,  non-vanishing of the
pair--wave function $\Delta$ is {\it not} a sufficient  signature of the
off--diagonal long--range order (ODLRO).
We can deduce this relationship from the definition of the
superconducting order parameter which implies,
\begin{eqnarray}
&&\Psi_{\ell\ell'}({\bf r}\tau,{\bf r}'\tau)\equiv
\langle\bar{c}_{\downarrow\ell}( {\bf r}\tau)\bar{c}_{\uparrow\ell'}( {\bf r'}\tau')\rangle
\nonumber\\
&&=\delta_{\ell\ell'}
\langle\bar{f}_{\downarrow\ell}( {\bf r}\tau)\bar{f}_{\uparrow\ell}( {\bf r'}\tau')e^{-i[\phi_\ell({\bf r}\tau)
-\phi_\ell({\bf r}'\tau')]}\rangle
\nonumber\\
&&\to\delta_{\ell\ell'}\Delta({\bf r}\tau,{\bf r}'\tau)\psi_0^2
\end{eqnarray}
where $\psi_0=\langle e^{i\phi_\ell({\bf r}\tau)}\rangle$.
The condensation of the ``flux-tubes" from the electron composite has a transparent
physical explanation.
The   phase factors  which are introduced into the hopping elements by the gauge 
transformation in Eq.(\ref{compo})  frustrate the motion in  the fermionic subsystem.
 However,  when charge  fluctuations become {\it phase coherent}, as  signalled by 
$\langle e^{i\phi_\ell({\bf r}\tau)}\rangle\neq 0$,
 the frustration of the kinetic energy is released. 
The role of the gap parameter $\Delta$ also
becomes  apparent: pairing among fermionic parts of the electron composites
is a necessary precondition for the existence of the microscopic
phase stiffnesses (c.f. Eq.(\ref{stiff})),
and, thereby, for the whole superconducting order. The opposite is obviously not true:
the pseudogap state with $\Delta\neq 0$ may be phase incoherent. 
However, the  appearance  of bulk phase coherence in the presence of large Coulomb 
interaction $U$, whose energy scale by far exceeds  that of microscopic phase stiffnesses, is
not so obvious. Therefore in the following
we elucidate the instrumental role of doping, since for the superconducting order to occur
the system  should be  brought in the  vicinity of the degeneracy point,
that is on the brink of change of the topological order.
%
\subsection{Effective NL$\sigma$M}
%
To proceed, we replace the phase degrees of freedom in Eq.(\ref{phasemodel}) by
the unimodular complex scalar $z_\ell({\bf r}\tau)=e^{i\phi_\ell({\bf r}\tau)}$ fields
via  suitable Fadeev-Popov resolution of unity
\begin{eqnarray}
1&\equiv&\int\left[{\cal D}^2 {z}\right]\prod_{{\bf r}\ell}
\delta\left(|{z_\ell({\bf r}\tau)}|^2-1\right)\times
\nonumber\\
&\times & \delta\left\{\Re\left[ z_\ell({\bf r}\tau)-e^{i\phi_\ell({\bf r}\tau)}
 \right]\right\}
\nonumber\\
&\times &\delta\left\{\Im\left[z_\ell({\bf r}\tau)-e^{i\phi_\ell({\bf r}\tau)} \right]\right\},
 \label{fp2}
\end{eqnarray}
where the  unimodularity constraint can be imposed using
Dirac $\delta$-functional, thus  bringing the partition function  into the following form:
\begin{equation}
{\cal Z}=\int\left[{\cal D}^2 {z}\right]\prod_{{\bf r}\ell}
\delta\left(|{z_\ell({\bf r}\tau)}|^2-1\right)e^{-{\cal S}[z,z^\star]}.
\label{zpsum}
\end{equation}
The (non-linear)  unimodularity constraint can be conveniently resolved with the help
of a real Lagrange multiplier  $\lambda$, so that the  phase action
in Eq.(\ref{phasemodel}) can be suitably expressed by the  effective non-linear $\sigma$-model
(NL$\sigma$M) represented by the action
\begin{widetext}
\begin{eqnarray}
{\cal S}[z,z^\star]&=&
\sum_\ell\int_0^\beta  d \tau\left\{-\sum_{{\bf r}\ell}\int_0^\beta   d\tau'
  {z}^\star_{\ell}({\bf r}\tau)\gamma^{-1}(\tau-\tau') {z}_{\ell}({\bf r}\tau)
-2{\cal J}_\|(\Delta)\sum_{{\langle{\bf r}{\bf r}'\rangle}}
\left[\frac{1}{2}{z}^\star_{\ell}({\bf r}\tau)
{z}_{\ell}({\bf r}'\tau)+c.c.\right]^2\right.
\nonumber\\
&-&{\cal J}'_\|(\Delta)\left.\sum_{{\langle\langle{\bf r}{\bf r}'\rangle\rangle}}
\left[\frac{1}{2}{z}_{\ell}^\star	({\bf r}\tau)
{z}_{\ell}({\bf r}'\tau)+ c.c.\right]
-2{\cal J}_\perp(\Delta)\sum_{{\bf r}}
\left[\frac{1}{2}{z}_{\ell}^\star({\bf r}\tau) 
{z}_{\ell+1}({\bf r}\tau)+c.c.\right]^2\right\},
\label{zaction}
\end{eqnarray}
\end{widetext}
where
\begin{eqnarray}
&&\gamma(\tau-\tau')=\frac{1}{Z_0}\int[{\cal D}\phi]
e^{i[\phi_\ell({\bf r}\tau)-\phi_\ell({\bf r}\tau')]}e^{-{\cal S}^{(0)}_{ph}[\phi]},
\nonumber\\
&& Z_0=\int[{\cal D}\phi]e^{-{\cal S}^{(0)}[\phi]}
\label{phasephase}
\end{eqnarray}
is the phase--phase  correlation function calculated with the action  in Eq.(\ref{freeaction})
(see  Appendix B).
Since a part of the action in Eq.(\ref{zaction}) is quartic in the unimodular 
$z$-fields we employ the mean-field like decoupling 
\begin{eqnarray}
&&\left[\frac{1}{2}{z}^\star_{\ell}({\bf r}\tau) {z}_{\ell'}({\bf r}'\tau)+c.c\right]^2\to 
\nonumber\\
&& \langle {z}^\star_{\ell}({\bf r}\tau) {z}_{\ell'}({\bf r}'\tau
\rangle\left[{z}^\star_{\ell}({\bf r}\tau) {z}_{\ell'}({\bf r}'\tau)+c.c\right]
\label{rule}
\end{eqnarray}
to perform the closed-form integration over $z$-variables in Eq.(\ref{zpsum}).
Here, the average $\langle\dots\rangle$ should be determined with  the
resulting effective action. To justify Eq.(\ref{rule}), we observe that,
formally, Eq.(\ref{rule}) follows from the decoupling of  quartic terms in
Eq.(\ref{zaction}) with the help of suitable Hubbard-Stratonovich transformation and
subsequent use of the saddle-point method with respect to the emerging
auxiliary variables. Next, we introduce the Fourier transformed variables
\begin{eqnarray}
{z}_{\ell}({\bf r}\tau)=\frac{1}{\beta NN_\perp}\sum_{\omega_n}\sum_{\bf q}{z}({\bf q},\omega_n)
e^{-i(\omega_n\tau-{\bf k}\cdot{\bf r}-k_zc\ell)},
\label{zfour}
\end{eqnarray}
where ${\bf q}\equiv({\bf k},k_z)$ with ${\bf q}=(k_x,k_y)$ and  $k_z$
labelinq  ``in-plane"  momenta and the wave vectors associated with the third
dimension along $c$-axis, respectively. With the help of Eq.(\ref{zfour})
the resulting quadratic action in the $z$-variables becomes
\begin{eqnarray}
	{\cal S}[z,z^\star]=
\frac{1}{\beta N N_\perp}\sum_{{\bf q}\omega_n}
  {z}^\star({\bf q},\omega_n)\Gamma^{-1}({\bf q},\omega_n)
 {z}({\bf q},\omega_n),
 \label{nlsmc}
\end{eqnarray}
where
\begin{equation}
\Gamma^{-1}({\bf q},\omega_n)=\lambda-{\cal J}({\bf q})+\gamma^{-1}(\omega_n)
\end{equation}
is the inverse propagator for the $z$-fields and 
\begin{eqnarray}
{\cal J}({\bf q})&=&\bar{\cal J}_\|(\Delta)\left[\cos(k_xa)
+\cos(k_ya)\right]
\nonumber\\
	&+&2\bar{\cal J}'_\|(\Delta)\cos(k_xa)\cos(k_ya)
\nonumber\\
&+&\bar{\cal J}_\perp(\Delta)\cos(k_zc)
\end{eqnarray}
is the dispersion associated with the microscopic phase stiffnesses, where
\begin{eqnarray}
&&\bar{\cal J}'_\|(\Delta)={\cal J}'_\|(\Delta)\langle {z}^\star_{\ell}({\bf r}\tau)
{z}_{\ell}({\bf r}+{\bf 1}_{st},\tau)\rangle
\nonumber\\
&&\bar{\cal J}_\perp(\Delta)={\cal J}_\perp(\Delta)\langle{z}_{\ell}^\star({\bf r}\tau) 
{z}_{\ell+1}({\bf r}\tau) \rangle.
\end{eqnarray}
Furthermore, $\gamma_{0}(\omega_n)$ is the Fourier transform of the
bare phase propagator  in Eq.(\ref{phasephase}), see Appendix B.
%
\subsection{SC critical boundary}
\begin{figure}
\begin{center}
\includegraphics[width=7cm]{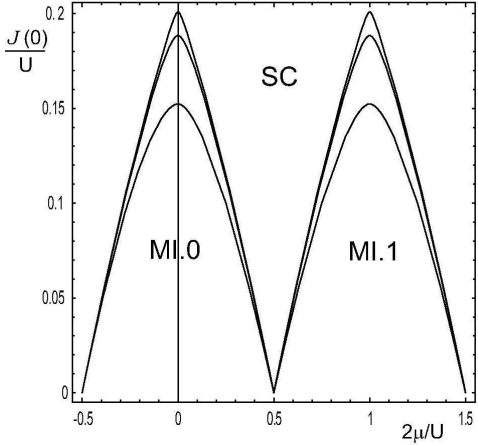}
\end{center}
 \caption{The ground state phase diagram resulting from the effective action in Eq.(\ref{phasemodel}).
 Here, 	in the filling-control transition, the control parameter is the chemical potential,
which is conjugate to the carrier density.
The  picture shows the arrangement  of Mott-insulating (incompressible) lobes MI.$m$ with
 topological order characterized but the winding number $m=0$ and $m=1$, respectively with the  phase coherent
superconducting ground state between and above them. For large Coulomb energy $U$
the phase coherent state is only possible in the vicinity of
the degeneracy point $2\mu/U=1/2$. The curves are plotted for different ratios of 
the inter- to intra-layer couplings as input parameters: ${\cal J}_\perp/{\cal J}_\|=0.001, 0.01$ and $0.1$ 
(from the top to the bottom) and show the proliferation of the superconducting state as the
stack of coupled two-dimensional planes  system crosses from 2D to 3D behavior.}
\label{fig3}
\end{figure}

%
At the critical boundary demarcating the long-range ordered phase-coherent 
true superconducting state the static and uniform ``order parameter" susceptibility
 diverges, so that the suitable  condition that can be read off from 
 the action in Eq.(\ref{nlsmc}) is
\begin{equation}
 \Gamma^{-1}({\bf 0},0)|_{\lambda=\lambda_c}=0 
 \label{lambdazero}
\end{equation}
which fixes the Lagrange parameter $\lambda$ at the transition boundary and  within
 the ordered state. The parameter $\lambda_c$ is given by the solution of the
 uni-modularity constraint equation:
\begin{equation}
1=\frac{1}{\beta N N_\perp}\sum_{{\bf q}\omega_n}
{ \Gamma}({\bf q},\omega_n).
\label{lambdaeq}
\end{equation}
The emerging ground-state phase diagram is depicted in Fig.\ref{fig3}.
It exhibits a periodic arrangement of phase incoherent  Mott-insulating lobes with the
superconducting state above and between them. Apparently,  at the degeneracy point
$\mu_{c}$ defined by 
\begin{equation}
\frac{2\mu_c}{U}=\frac{1}{2}
\end{equation}
the superconducting state is most robust.
Clearly, the above picture resembles that of a system  described by the well known
Bose-Hubbard Hamiltonian\cite{bosefisher} as a generic Hamiltonian for strongly correlated bosons.
It covers the physics originating from the competition between
the repulsive and kinetic term of the Hamiltonian,
whose magnitude are proportional, in the present setting,  to the parameters $U$
and phase stiffnesses, respectively. Moreover, this similarity is not accidental,
since this is just an example of statistical transmutation where bosons emerge as
fermions with attached ``flux tubes" as a result of the gauge transformation.\cite{wilczek2}

\begin{figure}
\begin{center}
\includegraphics[width=7cm]{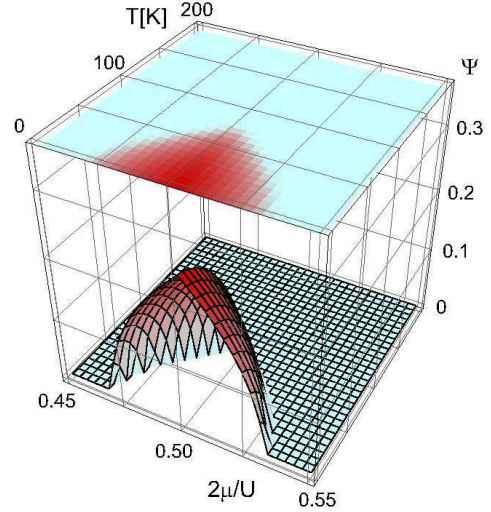}
\end{center}
 \caption{(Color online) Superconducting order parameter $\psi$ 
that signals the global phase stiffness shown  as a function of
the  temperature and chemical potential, for $U=4$eV, $J=0.15$eV , $t^\star=0.5$eV, 
$t'^\star/t^\star=0.25$ and  $t^\star_\perp=0.01$eV.
   }\label{fig4}
\end{figure}

Within  the phase coherent superconducting state order parameter is given by
\begin{equation}
1-\psi^2=\left.\frac{1}{\beta N N_\perp}\sum_{{\bf q}\omega_n}
{ \Gamma}({\bf q},\omega_n)\right|_{\lambda=\lambda_c}.
\end{equation}
With the aid of Eq.(\ref{propag}) and (\ref{gammaomega}), by performing the summation over Bose
Matsubara frequencies, we obtain 
\begin{widetext}
\begin{eqnarray}
1-\psi^2 &=&\frac{1}{4NN_\perp}\sum_{\bf q}
\frac{1}{\sqrt{
\frac{2\left[ {\cal J}(0) -{\cal J}({\bf q}) \right]}{U}+
h^2\left(\frac{2\mu}{U}  \right)}}
\left\{
\coth\left[\frac{\beta U}{4}\left(\sqrt{
\frac{2\left[ {\cal J}(0) -{\cal J}({\bf q}) \right]}{U}+ h^2\left(\frac{2\mu}{U}\right)
	}+h\left(\frac{2\mu}{U}\right) \right) \right]\right.
\nonumber\\
&+&\left.\coth\left[\frac{\beta U}{4}\left(\sqrt{
 \frac{2\left[ {\cal J}(0) -{\cal J}({\bf q}) \right]}{U} + h^2\left(\frac{2\mu}{U}\right)
	}-h\left(\frac{2\mu}{U}\right) \right) \right]  \right\}
	\label{opsi}
\end{eqnarray}
\end{widetext}
where $h(x-1/2)=x-[x]$, while $[x]$ is 
the greatest integer less than or equal to $x$. The remaining summation over the
momenta can be efficiently performed by resorting to the lattice density of
states as explained in  Appendix C.
The parameter $\psi$ as a function of temperature and
the chemical potential is depicted in Fig.\ref{fig4}, which shows substantial enhancement
of this quantity near the degeneracy point.
Note that  the three-dimensional
anisotropic lattice structure is essential, since  even a very small interplanar coupling 
renders the phase transition in the 3D universality
class as observed in cuprates.\cite{schneider} 
Thus, the absence of $t_\perp$ will suppress the bulk critical
temperature to zero, because  for  isolated stack of two-dimensional layers
the NL$\sigma$M strictly predicts $T_c=0$, in agreement with the
Mermin-Wagner theorem.

%
%
%
%
\section{Topological criticality  at the degeneracy point}
In the preceding paragraphs we  have shown that a theory of strongly
interacting electrons can be transformed to an equivalent description 
 of  weakly interacting fermions which are coupled to the ``fluxes" 
 of the  strongly fluctuating U(1) gauge field. 
In regard to the non-perturbative effects, we realized the  presence
of an additional parameter, the topological angle $\theta/2\pi \equiv2\mu/U$, 
 which  related to the chemical potential. We argued that the configuration space
 for the  phase field $\phi$ consists of distinct topological sectors,
 each characterized by an integer, entering the weight factors in the functional 
integral  and  counts the topological excitations of the system. On the other
 hand, the ground state degeneracy depends also on the topology of 
the configurational  space and  the  transition we encountered at 
$\theta/2\pi=1/2$ corresponds to an abrupt change of the ground state 
that is not related to any visible symmetry breaking. 
However, the existence of  different ground states that are  related to 
the topological properties of  the interacting electronic system 
is a hallmark of the topological order.\cite{wen}
The latter is not associated with the  symmetry breaking pattern, so  it cannot be 
characterized by  conventional order parameters in the  Landau sense.
We argue   that  the ground state degeneracy can be parametrized by a   topological
order parameter being the  average of the topological charge, i.e.,
the elements of the   homotopy group of the U(1) gauge group and this parameter
has a direct physical significance: in the large-$U$ limit the electron density
(i.e. the filing factor) is just given the mean topological charge rather than the
number of fermionic oscillators. Furthermore, in analogy to the Landau theory
where the divergence of the order parameter susceptibility signals the
phase transition between states with different symmetry, to indicate the change between
different topologically ordered states, we introduce the topological susceptibility
being a derivative of the mean topological charge with  respect to the statistical
angle $\theta$. Its divergence  is related
 to the existence of distinct ``vacua",  which cross in energy at the degeneracy point.
Subsequently, we show that the topological susceptibility has a  direct physical relevance, since
it is related to the charge compressibility. It  diverges at the degeneracy
point at $T=0$ and thus defines a novel type of topological  quantum criticality,
beyond the Landau paradigm of the symmetry breaking.
%
\subsection{Topological charge and the electron density}
\begin{figure}
\begin{center}
\includegraphics[width=6cm]{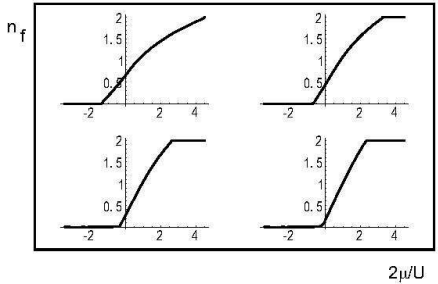}
\end{center}
 \caption{Evolution of the fermionic occupation number $n_f$ with increasing correlations at $T=0$
and for $t^\star=1$eV, $t^\star=0.5$eV, $t^\star=0.25$eV, and $t^\star=0.125$eV from the upper left
to the lower right. For all plots $U=4$eV, $J=0.15$eV and $t'^\star/t^\star=0.3$.
For large values of the Coulomb to band energy ratio the fermionic filling factor behaves as
$n_f\sim 2\mu/U$.
   }\label{fig5}
\end{figure}

\begin{figure}
\begin{center}
\includegraphics[width=6cm]{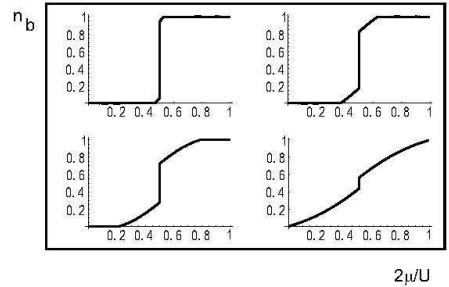}
\end{center}
 \caption{Evolution of the average topological number $n_b$ with decreasing correlations at $T=0$:
 $U=4$eV, $U=1$eV, $U=0.5$eV, and $U=0.2$eV from the upper left
 to the lower right. For all plots  $J=0.15$eV , $t^\star=0.5$eV, $t'^\star/t^\star=0.25$
and  $t^\star_\perp=0.01$eV.
For large values of the Coulomb to band energy ratio the electronic filling factor behaves as
$n_c\sim n_b$. The nearly linear dependence of $n_b$ as a function of the chemical
potential near the degeneracy point $2\mu/U=0.5$ signals the presence of the
global phase stiffness, see Eq.(\ref{fill}).
   }\label{fig6}
\end{figure}

\begin{figure}
\begin{center}
\includegraphics[width=7cm]{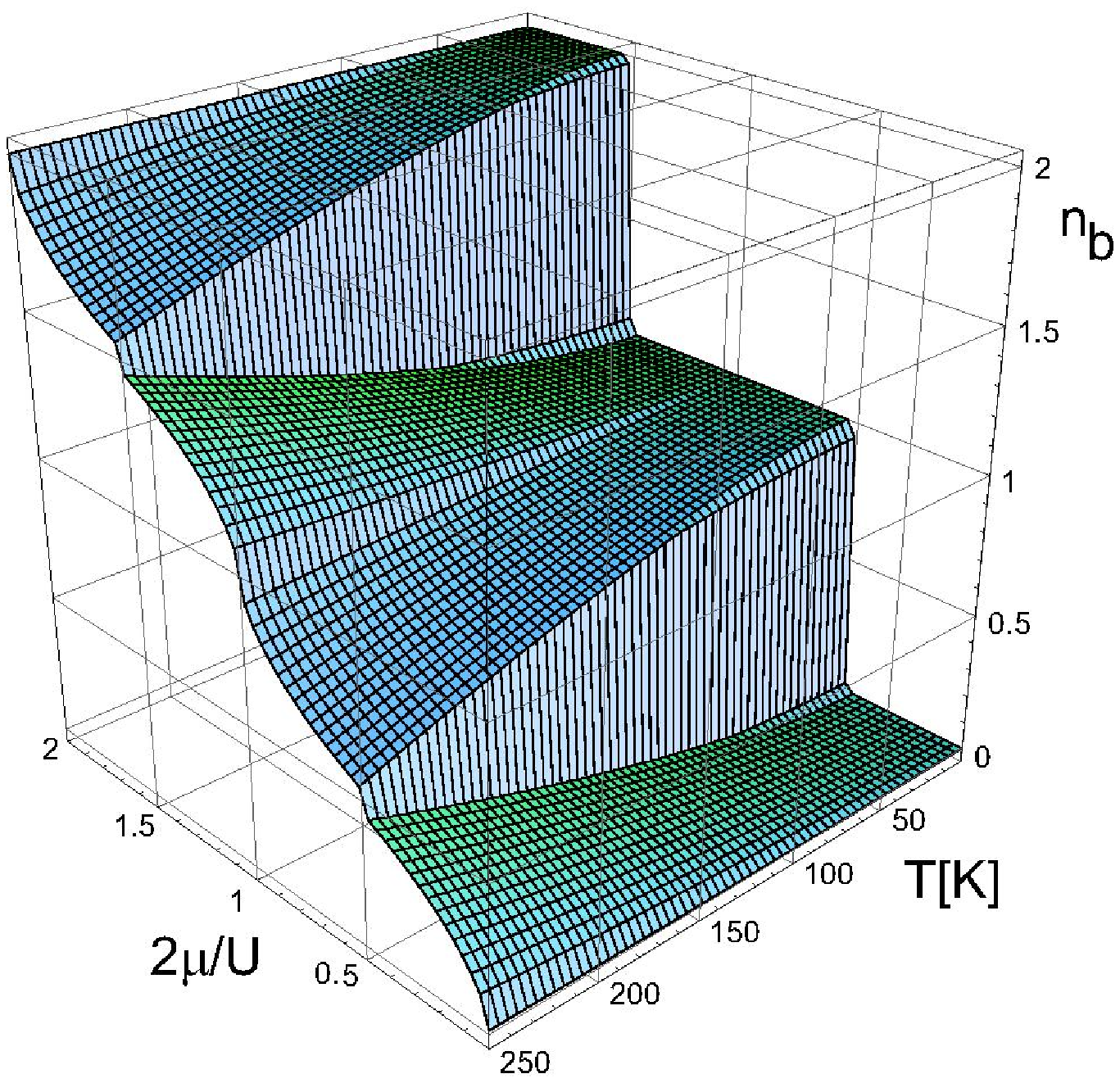}
\end{center}
 \caption{(Color online) Average topological number $n_b$ as a function of  temperature
 and chemical potential for$U=4eV$, $J=0.15$eV , $t^\star=0.5$eV, $t'^\star/t^\star=0.3$
and  $t^\star_\perp=0.01$eV.
   }\label{fig7}
\end{figure}

%
In addition to the Coulomb energy $U$ and temperature, the chemical potential $\mu$ plays
a crucial role in Mott transition, since it controls the
electron filling $n_e$.
An immediate implication of the composite nature of the electrons is that
the electron occupation number  (i.e., the average number of of electrons per site in
the Cu-O plane)
\begin{equation}
n_e=\frac{1}{N}\sum_{{\bf r}\alpha\ell}\left\langle
\bar{c}_{\alpha\ell}({\bf r}\tau)
c_{\alpha\ell}({\bf r}\tau)\right \rangle
\label{nenumb}
\end{equation}
consists of the fermion occupation coming from the fermionic part of the composite
and a topological contribution  resulting from the ``flux-tube" attachment:
\begin{eqnarray}
&&\left\langle \sum_\alpha\bar{c}_{\alpha\ell}({\bf r}\tau)
c_{\alpha\ell}({\bf r}\tau)\right \rangle=
\left\langle \sum_\alpha\bar{f}_{\alpha\ell}({\bf r}\tau)
f_{\alpha	\ell}({\bf r}\tau)\right \rangle
\nonumber\\
&+&
\frac{2}{iU}\left\langle\frac{\partial\phi_\ell({\bf r}\tau)}{\partial\tau}\right\rangle.
\label{cfphi}
\end{eqnarray}
The appearance of the topological contribution in Eq.(\ref{cfphi}) is not surprising given the fact that
``statistical angle", see Eq.(\ref{rhom}), depends on the chemical potential and the occupation number
is just its conjugate quantity. Owing that the
U(1) topological charge (the winding number) is given by
\begin{eqnarray}
m_\ell({\bf r})&=&\frac{1}{2\pi}\int_0^\beta
 d\tau\dot{\phi}_\ell({\bf r}\tau)
\nonumber\\
&=&\frac{1}{2\pi}\int_{
\phi_{\ell 0}({\bf r})}^{\phi_{\ell 0}({\bf r})+2\pi m_\ell({\bf r})}d\phi_\ell({\bf r}\tau)
\end{eqnarray}
the mean value of the density of the topological charge can be written after performing Legendre transformation 
as
\begin{eqnarray}
n_b&=&\frac{2\mu}{U}+\frac{2}{U}
\left\langle\frac{1}{i}\frac{\partial\phi_{\ell}({\bf r}\tau)}{\partial\tau}
   \right\rangle.
\end{eqnarray}
Therefore the  average electron occupation number $n_e$ is given by
\begin{eqnarray}
n_e=n_f+n_b -\frac{2\mu}{U}.
\label{nenbnf}
\end{eqnarray}
In the limit of strong (weak) correlations $n_e$ interpolates between topological $n_b$ (fermionic $n_f$)
occupation numbers. Clearly, in the large--$U$ limit  $\mu\to n_fU/2$, so that $n_e\to n_b$  and the system behaves 
as governed entirely by density of topological charge. The latter  behaves in the large $U$ limit 
as the typical density of hard-core bosons showing characteristic ``staircase" behavior, see Figs. \ref{fig5}-\ref{fig7}.
Indeed, in this limit the system is described by the quantum rotor action in Eq.(\ref{freeaction}), in which
the probability distribution function of the density of topological charge is gaussian and the problem 
 has single-site character  that can be analytically solved 
in a closed form:
\begin{eqnarray}
n_{0b}(\mu)&=&\frac{2\mu}{U} -\frac{1}{\beta}
\frac{\partial_\mu\theta_3\left[\frac{\beta\mu}{2\pi i}, 
e^{-{\beta U}/{4} }\right]}{\theta_3\left[\frac{\beta\mu}{2\pi i}, 
e^{-{\beta U}/{4} }\right]}
\end{eqnarray}
by making use of the Jacobi theta-function identity
\begin{eqnarray}
&&\frac{\partial_v\theta_3(iv,q)}{\theta_3(iv,q)}
=\sum_{m=1}^{\infty}\frac{(-1)^m4\pi iq^m}{1-q^{2m}}\sinh(2\pi m v).
 \end{eqnarray}
The calculation of the mean topological density for the interacting problem, i.e. with the full phase action
given by  Eq.(\ref{phasemodel}) involving phase stiffnesses is a bit demanding since spatial
correlations have to be included, as well. However, we can resort to the unimodular-filed 
NL$\sigma$M description given in Eq.(\ref{fp2}). The result for $n_b$ both within the Mott 
lobe and in the superconducting region is given by
\begin{eqnarray}
n_b=\left\{
	\begin{array}{l}
	n_b(\lambda),
\quad {\rm within\quad MI}\\
n_b(\lambda_0)-2\psi^2 h\left(\displaystyle\frac{2\mu}{U}\right),
\quad{\rm within\quad SC	}
	\end{array}\right.
	\label{fill}
\end{eqnarray}
where $\psi$ is the order parameter given by  Eq.(\ref{opsi}), while
\begin{widetext}
\begin{eqnarray}
n_b(\lambda) &=& n_{0b}(\mu)-\frac{1}{2NN_\perp}\sum_{\bf q}
\left\{
\coth\left[\frac{\beta U}{4}\left(\sqrt{
\frac{2\left[ {\cal J}(0) -{\cal J}({\bf q}) \right]}{U}+\delta\lambda+ h^2\left(\frac{2\mu}{U}\right)
	}+h\left(\frac{2\mu}{U}\right) \right) \right]\right.
\nonumber\\
&-&\left.\coth\left[\frac{\beta U}{4}\left(\sqrt{
 \frac{2\left[ {\cal J}(0) -{\cal J}({\bf q}) \right]}{U}+\delta\lambda + h^2\left(\frac{2\mu}{U}\right)
	}-h\left(\frac{2\mu}{U}\right) \right) \right]  \right\}.
\end{eqnarray}
\end{widetext}
with $\delta\lambda=\lambda-\lambda_0$. Here, the parameter $\lambda$ is self-consistently determined via
Eq.(\ref{lambdaeq}) whereas  $\lambda_0$ is given by the solution of Eq.(\ref{lambdazero}).
 The summation over the wave vectors
can be conveniently performed with the help of the lattice density of states.
%

%
\subsection{Topological susceptibility and  charge compressibility}
%
Mott insulators  have  a clear distinction from metals by vanishing 
of the charge compressibility at zero temperature, while this
quantity have a finite value in metals.\cite{imada}
In physical terms, the  charge compressibility measures 
the stiffness to the twist of the phase of the wave function in  the 
``imaginary time"  direction.
As we have shown, in the limit of strong correlations  the physical properties of the system 
are governed by the fluctuations of the topological charge.
Thus, the effects connected with the nontrivial topological configurations of the
gauge fields can be tested by performing 
the second derivative of the free energy with respect to the statistical
parameter $\theta$, see  Eq.(\ref{statangle}), which gives the topological
susceptibility,\cite{venez} i.e., the connected part of the two-point correlator of
the topological charge densities at zero momentum:
\begin{eqnarray}
\chi_t&=&\frac{1}{\cal Z}\sum_{m\in \pi_1(U(1))}\frac{\partial^2\rho(m)}{\partial\theta^2}{\cal Z}(m)
\nonumber\\
&-&\left[\frac{1}{\cal Z}\sum_{m\in \pi_1(U(1))}\frac{\partial^2\rho(m)}{\partial\theta^2}{\cal Z}(m)
\right]^2.
\label{zm12}
\end{eqnarray}
In Eq.(\ref{zm12}), as a result of the non-trivial topology in the group manifold group 
caused by the non-simply connected structure, the partition functions ${\cal Z}(m)$ are  given
 by  the functional integrals  taken over the field configurations in the topological class 
$m$ only. The full partition function ${\cal Z}$, as defined by Eq.(\ref{zm1}) involves all topological sectors.
Since the statistical angle parameter $\theta$ (and thereby the chemical potential $\mu$) 
acts as a ground state selector, the  topological susceptibility can  be conveniently
 employed to detect  transition between different topologically ordered states. Since these are labeled by the
average  topological charge, the abrupt change of this quantity will be signalled
by the divergence of $\chi_t$ at the degeneracy point
\begin{equation}
\lim_{\mu\to\mu_c}\chi_t(T=0,\mu)=\infty,
\end{equation}
where $\mu_c$ is the value of the chemical potential  at the degeneracy point.
The topological susceptibility in Eq.(\ref{zm12}) can be directly linked with
 the  physical quantities, namely the charge compressibility 
\begin{equation}
\kappa=\frac{\partial n_e}{\partial\mu},
\end{equation}
which expresses the total density response of the system
to a local change of the chemical potential.
It is related to the  shift of the electron chemical
potential  as a function of electron density which
can be measured e.g. through the shifts of spectral features in
photoemission spectra.\cite{chemical}
Taking the derivative  of Eq.(\ref{nenbnf}) with respect to the chemical potential we obtain
\begin{equation}
\frac{U}{2}\kappa=\frac{U}{2}\frac{\partial n_b}{\partial\mu}+\frac{U}{2}\frac{\partial n_f}{\partial\mu}-1.
\end{equation}
While the fermionic contribution $\partial n_f/\partial\mu$ is regular,
for the bosonic part $\partial n_b/\partial\mu$ one gets
\begin{equation}
2\pi\chi_t=\frac{U}{2}\frac{\partial n_b}{\partial\mu}.
\end{equation}
Therefore in the large $U$-limit  the charge compressibility is entirely governed by the
topological susceptibility and serves
to distinguish that
$\chi_t$ is zero in a Mott insulating\cite{mott} region while it
remains finite superfluid region and diverges at the degeneracy point.
%
\subsection{Electron mass enhancement at the degeneracy point}
\begin{figure}
\begin{center}
\includegraphics[width=6cm]{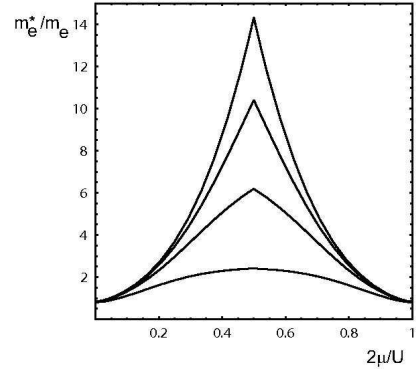}
\end{center}
 \caption{The ratio of the effective-to-bare electron mass parameter $m_e^\star/m_e$ as 
a function of the chemical potential$\mu$  in the vicinity of the degeneracy point for 
$U=4$eV, $J=0.15$eV , $t^\star=0.5$eV, $t'^\star/t^\star=0.25$, $t^\star_\perp=0.01$eV,
 and different temperatures $T=1$K, $50$K, $115$K and $300$K from the top to the
 bottom.
   }\label{fig8}
\end{figure}

\begin{figure}
\begin{center}
	\includegraphics[width=7cm]{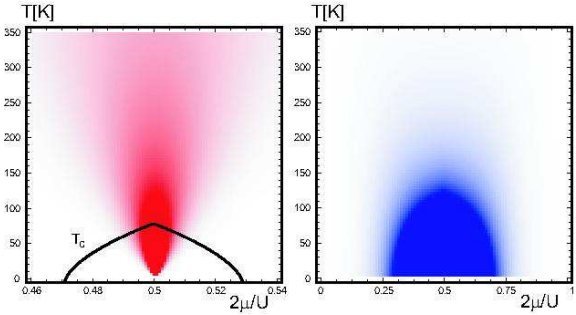}
\end{center}
 \caption{(Color online) Contrasting behaviors at the degeneracy
 point $T=0$, $2\mu/U=\frac{1}{2}$
and around it. Right picture: the density plot of the electron 
effective mass parameter $m_e^\star/m_e$ as a function
of temperature and the chemical potential showing  the strong depression 
of electron kinetic energies.
Left picture: The superconducting lobe with the density plot of the 
charge compressibility
$\kappa$ diverging at the degeneracy point. In the ``V-shaped" region 
 fanning out to finite temperatures
the electronic matter in its charge aspect  is very ``soft" 
( i.e. highly compressible as opposed to the Mott state).
 The values of the microscopic parameters
for  creation of the plots are the same as in Fig.\ref{fig3}.
   }\label{fig9}
\end{figure}

Another remarkable aspect of the transition from one topologically ordered
state to  another is the great enhancement of
the effective mass $m^\star_e$ of the electrons due to the
collapse of electron kinetic energies due to  the formation of the
degenerate state at $2\mu/U=0.5$. To estimate the change of $m^\star_e$
we calculate
\begin{eqnarray}
\frac{m^\star_e}{m_e}=\frac{\partial^2\epsilon_\|({\bf k})/\partial k_x^2|_{{\bf k}=0}}
{\partial^2\epsilon^\star_\|({\bf k})/\partial k_x^2|_{{\bf k}=0}}=\frac{1}{R},
\end{eqnarray}
where
\begin{eqnarray}
R=\left.\langle e^{-i[\phi_\ell({\bf r}\tau)
-\phi_\ell({\bf r}'\tau)]}\rangle\right|_{|{\bf r}-{\bf r}|=d}
\end{eqnarray}
is the band renormalization factor, see Eq.(\ref{baredress}),
where ${\bf d}$ stands for the lattice vector connecting nearest-neighbors sites
on a two-dimensional lattice. Figure \ref{fig8} illustrates the evolution
of the effective mass as a function of temperature in the vicinity of the
degeneracy point. Interestingly, at $2\mu/U=0.5$ which marks the ``topological
quantum critical point" the electronic matter in its charge aspect is very ``soft"
(see, Fig.\ref{fig9})
making  it very susceptible to transformation into alternative stable electronic
configurations, namely to superconductivity, which we are going to analyze.
%

%
\section{Phase diagram for cuprates}
\begin{figure}
\begin{center}
\includegraphics[width=7cm]{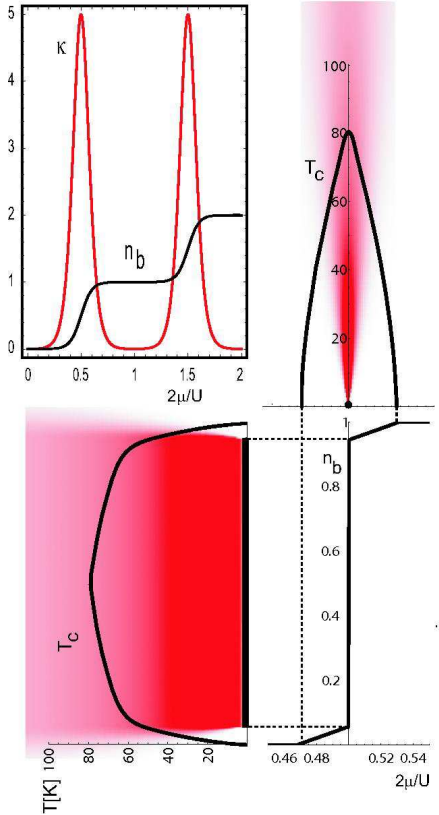}
\end{center}
 \caption{(Color online) Finite temperature phase diagram: 
the superconducting lobe $T_c(\xi)$ translated from the chemical 
potential ($\xi\equiv\mu$, upper right panel) to the
particle occupation number ($\xi\equiv n_b$, lower left panel). 
Shaded area: the density plot of the charge 
compressibility  $\tilde{\kappa}=U\kappa/2$. The  degeneracy point
($T=0,\frac{2\mu}{U}=\frac{1}{2}$), where
$\kappa$ diverges transforms into the critical line on the $n_b-T$ phase diagram.
Upper left panel: $\kappa$ and $n_b$ as a function of temperature for $T=0.1U$ 
showing the transition from the incompressible Mott state at $2\mu/U=1$ to
a highly compressible region around the degeneracy point.
The values of the  parameters for  creation of the plots are the same as in Fig.\ref{fig9}.
   }\label{fig10}
\end{figure}

There has been a considerable amount of controversy 
regarding the observed pseudogap phenomena in cuprates.\cite{tallon}
One general class of theories views the pseudogap phase as resulting from performed
pairs.\cite{emery,millis} The cuprates, however, are not in the strict  Bose condensation (``local pair")
limit, since the photoemission still reveals the presence of a large Fermi surface.
Furthermore, in the Bose limit, the chemical potential would actually be located beneath
the bottom of the energy band, which is also not the case.
The other class of scenarios	 (coined as ``competing order") consider the pseudogap
as not intrinsically related to superconductivity,
but rather proclaim it as  competitive with superconductivity. 
Most of these proposals involve either a charge
density wave\cite{zachar} or spin density wave\cite{dsc}, usually without long range order.
However, if there is a phase transition
underlying pseudogap formation, a  direct thermodynamic evidence
(i.e. non-analytic behavior of the specific heat, the susceptibility, or some
other correlation function of the system) must show up in existing experiments.
Unfortunately, none of the spectroscopic data support a picture where the pseudogap
phase represents a phase with true long range order. 
For the temperature-doping  phase diagrams the two delineated above 
scenarios generally predict that the pseudogap characteristic temperature
$T^\star(x)$ merges with $T_c(x)$ on the overdoped
side\cite{emery} (``precursor"  scenario) or
$T^\star(x)$ falls from a high value at low doping,
comparable to the exchange energy  to zero at a critical doping point
inside the superconducting dome\cite{qcp} (``competing order" picture).
Below we show that both scenarios are consistently accommodated within
the presence of topological order, degeneracy point and accompanying
phase coherence around it, as shown in Fig.\ref{fig10}.
We see the evolution of the charge compressibility $\kappa$ as a function of the chemical potential
  from the Mott insulator \cite{mott}
with $\kappa=0$ (at $2\mu/U=1$) to a point of degeneracy  on the brink of the
particle occupation change at $2\mu/U=1/2$ where $\kappa=\infty$ at $T=0$.
This is also the point on the phase diagram   from which the superconducting lobe emanates.
It is clear that, the nature of the divergence of $\kappa$ here has nothing to do
with singular fluctuations due to spontaneous symmetry breaking as in the ``conventional" phase transition.
Rather, this divergent response appears as a kind of topological protection
built in the system against
the small changes of $\mu$.
Further, $\kappa\to \infty$ implies that  $\partial\mu/\partial n_e$
becomes vanishingly small  at $T=0$ which results in the  chemical potential pinning,
 as observed in high-$T_c$ cuprates. \cite{chem1,chem2}

%
\subsection{Low energy scale pseudogap temperature $T^\star$}
%
In the pseudogap state at high temperatures one thus finds the coexistence of two distinct
components; a state with gaped fermionic excitations (described by the fermionic part of the composite electron)
and incoherent charge excitations (given by the attached ``flux-tube"),
 which, as the temperature is
lowered, enter the superconducting state. The
pseudogapped state is largely unaffected by the superconducting transition and
does not participate directly in the superconducting behavior.
As explained above the underlying  mechanism for the appearance of a
 gapped state with non-vanishing $\Delta$
is intimately connected with the antiferromagnetic
correlations represented by the AF exchange $J$.
We identify the temperature for which $\Delta$ sets in  with the pseudogap
temperature $T^\star$. This is also the temperature at which the microscopic 
phase stiffnesses in Eq.(\ref{stiff}) vanish,
so that
\begin{equation}
\left\langle e^{-i[\phi_\ell({\bf r}\tau)
-\phi_\ell({\bf r}'\tau)]}\right\rangle\to 0.
\label{renfactorzero}
\end{equation}
Using  results of Sec.VIA, the fermionic filling factor $n_f$ defined by Eq.(\ref{nfdef})
can be computed explicitly as
\begin{eqnarray}
n_f-1=-\frac{1}{N}\sum_{\bf k}
\left[\frac{\epsilon^\star({\bf k})-\bar{\mu}}{{ E}({\bf k})}\right]
\tanh\left[
\frac{\beta{E}({\bf k})}{2} \right].
\end{eqnarray}
With the help of Eq.(\ref{renfactorzero}) and using Eq.(\ref{rvbsol}) and we obtain
that at $T^\star$
\begin{eqnarray}
&&\frac{1}{J}=\frac{1}{2|\bar{\mu}|}\tanh\left(\frac{\beta|\bar{\mu}|}{2}  \right),
\nonumber\\
&&n_f-1=\frac{\bar{\mu}}{|\bar{\mu}|}\tanh\left(\frac{\beta|\bar{\mu}|}{2}  \right).
\label{tstareq}
\end{eqnarray}
By eliminating the chemical potential from Eq.(\ref{tstareq}) we get
\begin{equation}
 \frac{k_BT^\star}{J}={{\,\left| {n_f}-1\right| }\over{\displaystyle\ln\left[-{{2\,%
 \left| { n_f}-1\right| +(n_f -1)^2  +1 }\over{2\,%
 \left| {n_f}-1\right| -(n_f-1)^2-1}}\right]}}
 \label{tstar}
\end{equation}
revealing that $T^\star$ is a universal function of the fermionic filing number $n_f$.
Approaching half-filing ($n_f=1$) we can infer from Eq.(\ref{tstar})
\begin{equation}
\lim_{n_f\to1}{k_BT^\star(n_f)} =\frac{J}{4\ln(3)}\approx 0.228 J\equiv k_BT^\star_{max},
\end{equation}
where $T^\star_{max}$ is the maximum value of the pseudogap temperature $T^\star$.
For $J=0.15$eV we obtain $T^\star_{max}=396K$, see Fig.\ref{fig12}.

%
\subsection{Effect of doping on  AF exchange and $x-T$ phase diagram for cuprates}
\begin{figure}
\begin{center}
\includegraphics[width=6cm]{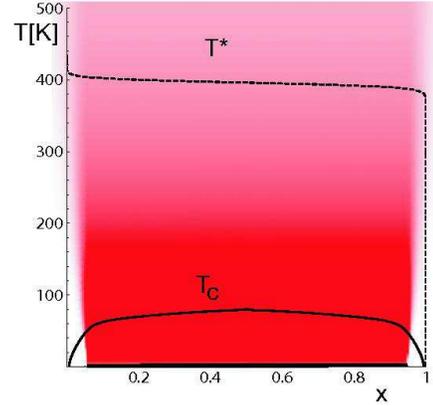}
\end{center}
 \caption{(Color online) Temperature--doping $(x=1-n_b)$ phase diagram calculated  for 
 the doping independent  AF exchange $J$ (for the values of the parameters, see Fig.\ref{fig9}).
 Depicted is the superconducting lobe (bounded by the critical temperature $T_c(x)$)
 and the pseudogap temperature $T^\star$ which marks the region of non-vanishing
 microscopic phase stiffnesses (below $T^*(x)$). Shaded area: intensity plot of the
 charge compressibility $\kappa$ diverging along the critical $T=0$ line (see, previous figure).
   }\label{fig11}
\end{figure}

%
It is well known that the  antiferromagnetic exchange $J$ originates from the interplay between
on--site repulsion ($U$) and the delocalization energy ($t$). 
The effect can be derived straightforwardly by expanding the energy to the
second order in the hopping matrix element. This involves virtual
double occupation and can be represented  by an exchange process taking place on neighboring lattice
sites:
\begin{eqnarray}
&&|\uparrow,\downarrow\rangle\stackrel{t}\rightarrow
\overbrace{|\uparrow\downarrow,0\rangle}^{1/U}
\stackrel{t}\rightarrow|\downarrow,\uparrow\rangle\nonumber\\
&&|\uparrow,\downarrow\rangle\stackrel{t}\rightarrow
\underbrace{|0,\uparrow\downarrow\rangle}_{1/U}
\stackrel{t}\rightarrow|\uparrow,\downarrow\rangle
\label{exproc}
\end{eqnarray}
which yields a contribution $\sim t^2/U$. There are also processes that are prohibited
by the Pauli exclusion principle such as $|\uparrow,\uparrow\rangle\stackrel{t}\rightarrow 0$.
However, with increasing hole doping (i.e., departing from the half-filling) a given electron has fewer neighboring electrons to pair with, which results in degradation of the  exchange process
described in Eq.(\ref{exproc}). 
Therefore, the AF exchange is leading to an effective interaction,\cite{effJ} which is a steadily decreasing function of  $x$, vanishing at the critical doping $x_c$ :
\begin{equation}
J_{\rm eff}(x) = J(1 - Kx),
\label{jeff}
\end{equation}
where $K$ is treated as the lattice connectivity
and the factor of $K=4$ is the coordination
number on the 2D square lattice. 
Therefore, the vanishing of $J_{\rm eff}$ determines the  critical doping value $x_c = 0.25$.
It is clear by inspecting Eq.(\ref{rvbsol}) that diminishing of the AF exchange  with doping will 
spoil the RVB pairing of the fermionic part of the electron composite by suppressing the $d$-wave gap 
function $\Delta({\bf k})$. Since phase stiffnesses in Eq.(\ref{stiff}) rest on $\Delta({\bf k})$ 
the doping dependence of $J_{\rm eff}(x)$ can be directly translated into the calculation of the
$x-T$ phase  diagram that involves the doping effect on $J_{\rm eff}$, see Fig.\ref{fig12}.
By comparing Fig.\ref{fig11} and Fig.\ref{fig12} we can  clearly see that the
diminishing of the superconductivity  in the overdoped region is just the result of the
pair-breaking effect triggered by the doping dependent AF exchange.
To summarize: $T^\star(x)$ demarcates the region  of non-zero microscopic
phase stiffnesses  which persist in the region characterized by the non-vanishing
of the spin gap, as observed in high-frequency conductivity
measurements.\cite{corson} The origin of the spin gap is purely
electronic and results from the restricted space of available
states that strongly correlated excitations on neighboring
sites encounter. It is described by the resonating valence
bond singlet $d$-wave spin pairing of the  fermionic part of the electron composite and
is controlled by the AF super-exchange parameter $J$.
The coincidence of the Fermi surface with
the minimum gap locus as obtained from  ARPES  measurements\cite{arpes3} also supports of a
pairing gap interpretation of the pseudogap.\cite{rand}
%
\subsection{Crossover to the ``strange metal" state at $T_g$: high energy scale feature}
\begin{figure}
\begin{center}
\includegraphics[width=6cm]{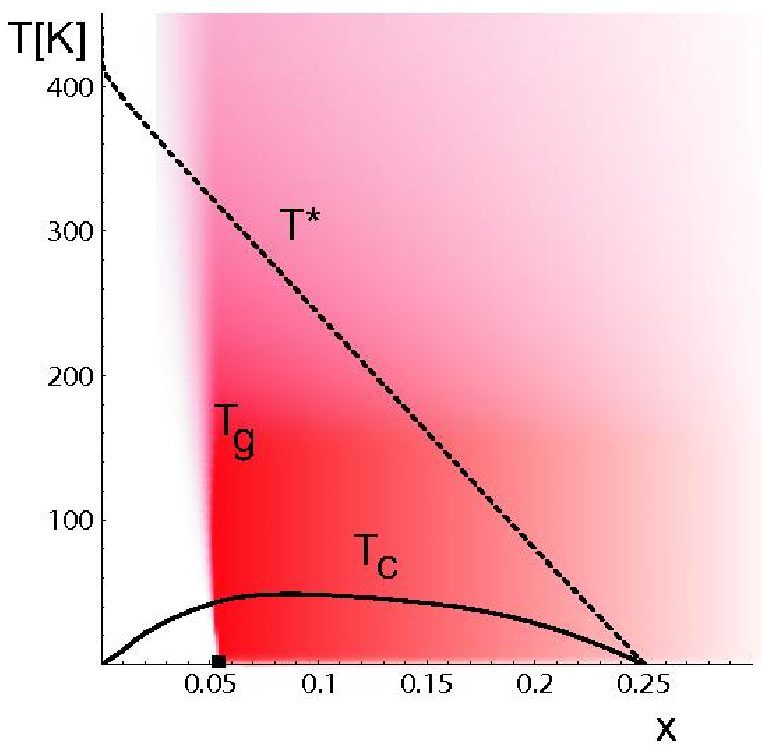}
\end{center}
\caption{
Temperature--doping  phase diagram as in Fig.\ref{fig11} but
with doping dependent  AF exchange parameter 
$J_{\rm eff}(x)$ given by Eq.(\ref{jeff}) with $K=0.25$. The temperature
$T_g(x)$ marks the sudden onset  of greatly enhanced charge compressibility $\kappa$
in the ``strange metal" state	
as departing from the half-filling ($x=0$).  Note, that the characteristic energy 
scale of these charge fluctuations (controlled by the Mott energy $U$ ) is much
larger than that of  spin gap, controlled by  AF exchange $J$.}
\label{fig12}
\end{figure}

%
The doping-dependent characteristic
temperature $T^\star$  in Eq.(\ref{tstar}) at which this pseudogap opens
is in the underdoped region significantly larger than $T_c$.
The physical reason for this is transparent: $T^\star$ marks the
region of non-vanishing phase stiffness, albeit without global
phase coherence (that appears at much lower temperature $T_c$).
However, when the copper oxide superconductor is
driven in the normal state by applying a high magnetic field, a
clear pseudogap feature at a similar energy scale to the superconducting
gap is observed in the quasiparticle tunnelling spectra and
the pseudogap feature persists up to the
highest applied fields and does not depend on the magnetic field.\cite{bobin}
Surprisingly, the Hall coefficient does not vary monotonically with doping but rather
exhibits a sharp change at the optimal doping level for superconductivity.\cite{hall}
This observation would support the idea that two competing
ground states underlie the high-temperature superconducting phase.
From this perspective one has to conclude that  any prospective  order 
consistent with these observations implies that the pseudogap coexists
with superconductivity and is
essentially unchanged by a large applied external field.
It is clear that sudden onset of the pseudogap
at critical doping right at the point where the
rigidity of the condensate wave function is at its
maximum would be very difficult to
reconcile with the precursor scenario,
but is very consistent with the onset of correlations which compete
with superconductivity. This is precisely the outcome of the
topological order which differentiates the electronic ground state
into two states  labeled by the topological winding number
and with the degeneracy point separating  them. It
controls a remarkable concurrence between normal state
properties and the ground-state superconducting
order  setting up a unique critical doping
point in the phase diagram where the transport properties change
very suddenly and where superconductivity is most robust.
Therefore we identify the
crossover line  $T_g(x)$  where the charge compressibility  undergoes a sudden change
as an  additional  boundary  hidden in the cuprate phase diagram, see Fig.\ref{fig12}.
It is important to realize that in contrast to $T^\star(x)$ the crossover line
$T_g(x)$ is controlled by the highest energy scale in the problem, namely the Mott scale
set up by $U$. This explains why the anomalous behavior still persists even the
superconducting order is suppressed, eg. by the strong magnetic field.
Considering  the Hall effect one has to
conclude that the mobile carrier density varies considerably
with both doping and temperature. On the other hand, the
ARPES Fermi surface remain essentially unchanged in the
whole of the metallic doping range suggesting a constant
density of states. These apparently contradictory results could
be reconciled by observing that ARPES is sensitive to the
momentum-space occupation and therefore detect the excitation described by the
fermionic part of the electron composite, whereas the charge transport properties
are governed mainly by the ``flux tubes" which constitute   charge   collective variables.
Given fact that  the inverse of the Hall coefficient is proportional to the number of carriers
$1/R_H\sim n_e$ and $n_e$ is governed by the topological charge $n_b$ it is apparent 
why $1/R_H$ jumps in the vicinity of the QCP that is of topological origin.
%

\section{Quantum  protectorate}
It is often difficult to formulate a fully consistent and adequate microscopic theory of
complex cooperative phenomena and 
great advances in the solid-state physics  are
to a great  extent due to the use of	 simplified and schematic
model representations for the theoretical interpretation.
In particular, the method of model
Hamiltonians has proved to be very effective.
However, as it was recently argued\cite{pl}, ab-initio
computations have failed completely to explain
phenomenology of high $T_c$ cuprates: it would appear
that not only the deduction from microscopics
failed to explain the wealth of crossover
behaviors, but as a matter of principle it probably cannot
explain it. Therefore it was  concluded that a more appropriate starting point
would be to focus on the results of experiments in the hope of identifying 
the corresponding ``quantum protectorates" –- stable states of matter whose generic low energy
properties, insensitive to microscopics, are determined by a ``higher organizing principle and nothing else".
From this perspective, each protectorate can
be characterized by a small number of parameters, which can be determined experimentally,
but which are, in general, {\it impossible}		 to calculate
from first principles.\cite{pl}
However, as we saw, 
a system with many microscopic degrees of freedom
can have ground states whose degeneracy is determined by
the topology of the system. 
Prototypes of this
kind of systems are provided by fractional quantum Hall
effect.
For example, the ground state degeneracy in FQH liquids is
not a consequence of symmetry of the Hamiltonian. It
is robust against arbitrary perturbations, even impurities
that break  the symmetries in the Hamiltonian.
Thus the topological ground state degeneracy on non-trivial
manifolds provides a precise theoretical distinction
between a topological and conventional order.
The Hilbert space of quantum states decomposes into distinct topological 
sectors, each sector remaining isolated under the action of
local perturbations. 
This is a signature of its topological nature.
Choosing the  states from ground
states in different sectors protects these states from unwanted
mixing through the change of system parameters- protection  within the sector
is secured through a gapped excitation spectrum.
In particular, we found that
for strong correlations the system  is governed
by the topological Chern numbers. However, the Chern number
 is a topologically conserved quantity
and is ``protected"  against
the small changes of system parameters. Being 	an integer it can not change 
at all if it has to change continuously.
However, changing the interaction by a large amount may cause
abrupt changes in ground state properties
described by a different topological quantum number, which
leads to a change of topological order. This kind of stability  might be generic
for quantum  systems governed by topologically non-trivial   groups manifolds. 
Therefore one is left not only with  the low-energy
principle (the classic prototype being the Landau Fermi liquid),
but the emergent physical phenomena are regulated  also by  topological
principles that have a property of their insensitivity to
microscopics
and this quantum protectorate functions  under certain topological
environments, through conserving  of 
topological charges.
%
\section{Summary and discussion}
%
In the present work focusing on the $t-t'-t_\perp-U-J$ 
model it is shown that  the topological excitations of charge  given by the collective U(1) phase field 
in a form of ``flux tubes"  attached to fermions
can reproduce many robust features present in the phase diagram of high-$T_c$ cuprates,
thus substantiating one of the emerging paradigms in the condensed matter
physics, namely  the ubiquitous competitions in strongly correlated systems.
The fundamental entities that carry  charge (and spin) in the copper oxides are
no longer the  usual Landau quasiparticles
but  the ``flux tube" fermion composites, so that  the charge  is no longer tied to the 
Fermi statistics. When charges are  ``liberated" then they can  condense leading to superconductivity.
This picture naturally leads to the pseudogap physics that is observed in the underdoped
cuprates, which  originates from   the momentum pairing (in a $d-$wave pattern) of the
fermionic parts of the electron  composite controlled by the antiferromagnetic superexchange $J$. 
This underlines the necessity of the fundamental  concept of  fermion pairing in achieving 
the superconductivity.
In the mathematical structure of the theory the gauge field is governed
by the U(1) Chern-Simons term in the action of
 purely topological  nature.  From the canonical point of view
the Hilbert space of a quantum theory has a non-trivial structure
marked by the topological sectors which corresponds to
 a set of degenerate ground states.
The topological ground state degeneracy  provides a precise theoretical distinction
between a topological and conventional order:
states with the same conventional broken symmetries
may still be distinguished from each other on the basis of whether
they are characterized by different topological quantum numbers
captured by the homotopy theory gauge group manifold.
In this paper, it has been shown that physics of the Mott transition is successfully covered in
a topological framework. The natural order parameter for the Mott
transition is the  topological charge related to
electron  concentration for the filling-control scenario
that selects topologically ordered states.
Expectation value of the density of topological charge,
determines  also a dominant contribution to the
topological susceptibility which, in turn, 
is related to the charge compressibility of the system. It  diverges at the degeneracy
point at zero-temperature and defines a novel type of topological  quantum criticality,
beyond the Landau paradigm of the symmetry breaking.
Although the charge compressibility is completely suppressed in the Mott insulator because
of the Mott gap, the criticality on the ``strange metallic" side can be
described adequately by the  divergence of the charge compressibility  at zero temperature. 
 This critical enhancement of the density fluctuations extends to finite-temperature and is controlled by the
Mott energy scale. This gives rise to  another
crossover line hidden in the cuprate phase diagram, where the charge compressibility 
 undergoes a sudden change to the ``strange metal" state.
The crossover is governed  by the Coulomb energy $U$,
so that the  density fluctuations at the  instability towards
the superconductivity  surpass the  effects of the spin
fluctuation mechanisms  governed by the antiferromagnetic exchange 
extensively studied for the cuprate superconductors. 
This  clearly demonstrates the inseparability of the high-energy Mott scale 
from the low-energy physics in the cuprate problem  and redefines the role of
the chemical potential  from a quantity  that simply demarcates the boundary between filled
and empty states to a selector of topologically ordered electronic ground state.

\appendix
%
\section{Microscopic phase stiffnesses}
%
The microscopic {phase stiffnesses} to the lowest order in the hopping amplitudes
are given by
\begin{eqnarray}
{\cal J}_\|(\Delta)&=&
 \frac{2t^2}{N\beta}\sum_{{\bf k}\nu_n}{\cal F}^\star({\bf k}\nu_n)
 {\cal F}({\bf k}\nu_n)
\nonumber\\
{\cal J}'_\|(\Delta)&=&\frac{t'}{\beta N}\sum_{{\bf k}\nu_n}
\cos(ak_x)\cos(ak_y){\cal G}({\bf k}\nu_n)
\nonumber\\
{\cal J}_\perp(\Delta)&=&\frac{1}{N}
\sum_{\bf q}\frac{t^2_\perp({\bf q})}{\beta N}\sum_{{\bf k}\nu_n}{\cal F}^\star({\bf k}\nu_n)
{\cal F}({\bf k}\nu_n).
\label{stif1}
\end{eqnarray}
Explicitly, after performing frequency and momentum sums in Eq.(\ref{stif1}) we obtain
\begin{eqnarray}
{\cal J}_\|(\Delta)&=&\frac{t^2}{4}
\int_{-2}^{2}dxdy\frac{x^2y^2}{y^2-x^2}\rho(x)\rho(y)\times
\nonumber\\
&&\times
\left\{\frac{\tanh\left[\frac{1}{2}\beta \epsilon(x)  \right]}{\epsilon(x)}-
\frac{\tanh\left[\frac{1}{2}\beta \epsilon(y)  \right]}{\epsilon(y)}\right\},
\nonumber\\
{\cal J}'_\|(\Delta)&=&-t'\bar{\mu}\int_{-2}^2 dx
\frac{\bar{\rho}(x)}{\epsilon(x)}
\tanh\left[\displaystyle
\frac{1}{2} \beta\epsilon(x)  \right],
\nonumber\\
{\cal J}_\perp(\Delta)&=&{\frac{9{t}^2_{\perp}|\Delta|^2}{16}}
\int_{-2}^2dx 
\frac{x^2\rho(x)}{\epsilon^{3}(x)}
\left\{2\tanh\left[\frac{\beta \epsilon(x)}{2}  \right]
\right.
\nonumber\\
&&-\left.\beta\epsilon(x)
{\rm sech}^2\left[ \frac{\beta \epsilon(x)}{2}  \right]  \right\}.
\label{stiff}
\end{eqnarray}
Here,  $\epsilon(x)=\sqrt{ \bar{\mu}^2+|\Delta|^2 x^2}$ and
\begin{eqnarray}
&&\rho(x)=({1}/{\pi^2}){\bf K}(\sqrt{1-({x^2}/{4})} )
\nonumber\\
&&\bar{\rho}(x)=\rho(x)-
(2/\pi^2){\bf E}( \sqrt{1-({x^2}/{4})})
\end{eqnarray}
where ${\bf K}(x)$ and  ${\bf E}(x)$ are the complete  elliptic integrals 
of the first and second kind, respectively.\cite{EllipticFunction}

%
\section{Phase-phase correlation function}
%
By performing the functional integration
over the phase variables  in Eq.(\ref{phasephase}) we obtain
\begin{eqnarray}
&&\gamma^{-1}(\tau-\tau')=
\frac{\vartheta_3\left(\frac{2\pi\mu}{U}+\pi\frac{\tau-\tau'}{\beta},
e^{-\frac{4\pi^2}{\beta U}}  \right)}{\vartheta_3\left(\frac{2\pi\mu}{U},
e^{-\frac{4\pi^2}{\beta U}}  \right)}
\nonumber\\
&&\times\exp\left\{-\frac{U}{4}\left[|\tau-\tau'|-\frac{(\tau-\tau')^2}{\beta} \right]  \right\},
\label{propag}
\end{eqnarray}
where $\vartheta_3(z,q)$ is the Jacobi theta function,\cite{EllipticFunction}
which comes from the topological part of the functional integral over the
phase variables. The function $\vartheta_3(z,q)$ is defined by
\begin{equation}
\vartheta_3(z,q)=1+2\sum_{n=1}^\infty\cos(2nz)q^{n^2}
\end{equation}
and is $\beta$-periodic in the
"imaginary-time" $\tau$ as well as in the variable $2\mu/U$ with the period of unity.
Fourier transforming one obtains
\begin{eqnarray}
\gamma(\omega_n)=\frac{1}{Z_0}\sum_{m=-\infty}^{+\infty}
\frac{\frac{8}{U}\exp\left[-\frac{\beta U}{4}\left(m-\frac{2\mu}{U}\right)^2  \right]}
{ 1-4	\left[\left(m-\frac{2\mu}{U} \right)
-\frac{2i\omega_n }{U} \right]^2},
\label{gammaomega}
\end{eqnarray}
where
\begin{equation}
{\cal Z}_0=\exp\left(-\frac{\beta \mu^2}{U}\right)\theta_3\left(\frac{\beta\mu}{2\pi i}, 
e^{-{\beta U}/{4} }\right)
\end{equation}
is the partition function for the ``free" rotor Hamiltonian in Eq.(\ref{rot}).
%
%
\section{Lattice density of states}
%
In this Appendix we give the explicit formulas for the
densities of states (DOS) for the anizotropic  three-dimensional lattice 
that is helpful for evaluation of the  sums  over the momenta 
that appear in Section VII and VIII of  the  present paper.
Our starting point is the dispersion relevant for the two-dimensional lattice with next-nearest interactions:
\begin{eqnarray}
E({\bf k})=&-&2t\cos(ak_x)-2t\cos(bk_y)
\nonumber\\
&+&4t'\cos(ak_x)\cos(ak_y)=
\nonumber\\
&=&{2t}\left[ -\cos(ak_x)-\cos(ak_y)\right.
\nonumber\\
&+&r\left.\cos(ak_x)\cos(ak_y) \right].
\end{eqnarray}
The choice of such a dispersion is obviously motivated by its relevance
 as a simple means of modelling the quasi-particle band in
 the high-$T_c$ cuprates. The density of states reads:
\begin{eqnarray}
\rho(E)=&&\int_{-\pi/a}^{\pi/a}\frac{dk_x}{(2\pi/a)}
\int_{-\pi/a}^{\pi/a}\frac{dk_y}{(2\pi/a)}\times
\nonumber\\
&&\times\delta[E-E({\bf k})]\equiv\frac{1}{2t}\tilde{\rho}(\epsilon)
\label{dos2d}
\end{eqnarray}
with
\begin{eqnarray}
\tilde{\rho}(\epsilon)&=&\frac{{\bf K}\left[
\sqrt{\frac{ 4-(\epsilon-r)^2 }{4(1+r\epsilon)} }\right]}{\pi^2 \sqrt{1+r\epsilon}}
\left[\Theta\left(2+r-\epsilon	\right)\times\right.
\nonumber\\
&&\left.\Theta\left(\epsilon+r\right)+\Theta\left(-r-\epsilon\right)\Theta\left(\epsilon +2-r\right)\right],
\label{dosyyy}
\end{eqnarray}
where $\epsilon=E/2t$ and $r=2t'/t$. Here $\Theta(x)$ is the Heavyside (unit-step) function.
The expression in Eq.(\ref{dosyyy}) is valid 
for $r\le 1$. For $r>1$ one has to make the replacement $\tilde{\rho}(\epsilon)\to |\Re e\tilde{\rho}(\epsilon)|$.

The effect of inter-planar ($c$-axis) interaction can be incorporated as well via the following dispersion
\begin{eqnarray}
E_{3d}({\bf k})=E({\bf k})-2t_z\cos(cq_z).
\end{eqnarray}
In the presence of $t_z$ the system is a tree-dimensional anisotropic one for which the density of states
becomes
\begin{eqnarray}
\rho_{3d}(E)=&&\int_{-\pi/a}^{\pi/a}\frac{dk_x}{(2\pi/a)}\int_{-\pi/b}^{\pi/a}\frac{dk_y}{(2\pi/a)}
\int_{-\pi/c}^{\pi/c}\frac{dq_z}{(2\pi/c)}\times
\nonumber\\
&&\times\delta[E-E_{3d}({\bf k})]\equiv\frac{1}{2t}\tilde{\rho}_{3d}(\epsilon).
\end{eqnarray}
Performing the integration over $k_z$ we obtain
\begin{eqnarray}
\rho_{3d}(E)=\frac{1}{2\pi t} \int_{-\infty}^{+\infty} d\xi\frac{\tilde{\rho}(\xi)
\Theta\left[r^2_z-(\epsilon-\xi )^2 \right]}
{\sqrt{r^2_z-(\epsilon-\xi )^2}}
\label{dosxx}
\end{eqnarray}
in a form of the convolution involving previously calculated DOS in Eq.(\ref{dosyyy}).


 \end{document}